\documentclass{aa}  
\usepackage[normalem]{ulem}
\usepackage{graphicx}
\usepackage{color}
\usepackage{xcolor}
\usepackage{hyperref}
\usepackage{txfonts}
\begin{document} 

   \title{Simulating the LOcal Web (SLOW)}

   \subtitle{I. Anomalies in the local density field}

   \author{Klaus Dolag\inst{1,2}
          \and
          Jenny G. Sorce\inst{3,4,5}
          \and
          Sergey Pilipenko\inst{6}
          \and
          Elena Hern\'andez-Mart\'inez\inst{1}
          \and
          Milena Valentini\inst{1}
          \and
          Stefan Gottl{\"o}ber\inst{5}
          \and
          Nabila Aghanim\inst{4}
          \and
          Ildar Khabibullin\inst{1,2}
          }

   \institute{Universit\"ats-Sternwarte, Fakult\"at für Physik, Ludwig-Maximilians-Universit\"at M\"unchen, Scheinerstr. 1, 81679 M\"unchen, Germany
         \and
   Max-Planck-Institut für Astrophysik, Karl-Schwarzschild-Straße 1, 85741 Garching, Germany
         \and
   Univ. Lille, CNRS, Centrale Lille, UMR 9189 CRIStAL, F-59000 Lille, France
         \and
   Universit\'e Paris-Saclay, CNRS, Institut d'Astrophysique Spatiale, 91405, Orsay, France   
         \and
   Leibniz-Institut f\"{u}r Astrophysik (AIP), An der Sternwarte 16, 14482 Potsdam, Germany 
         \and
   P.N. Lebedev Physical Institute of the Russian Academy of Sciences, Profsojuznaja 84/32 Moscow, Russia, 117997
             }

   \date{Received December 1, 2022; submitted}

% \abstract{}{}{}{}{} 
% 5 {} token are mandatory
 
  \abstract
  % context heading (optional)
  % {} leave it empty if necessary  
   {Several observations of the local Universe point towards the existence of very prominent structures. The presence of massive galaxy clusters and local super clusters on the one hand, but also large local voids and under-densities on the other hand. However, it is highly non trivial to connect such different observational selected tracers to the underlying dark matter (DM) distribution.}
  % aims heading (mandatory)
   {Therefore, it is needed to construct mock catalogues of such observable tracers using cosmological hydro-dynamical simulations. Such simulations have to follow galaxy formation physics on the one hand and on the other hand have to be constrained to reproduce the local Universe. Such constraints should be based on observables which directly probe the full underlying gravitational field, like the observed peculiar velocity field, to provide an independent test on the robustness of those distinctive structures.}
  % methods heading (mandatory)
   {We used a 500 $h^{-1}$Mpc large constrained simulation of the local Universe to investigate the anomalies in the local density field as found in observations. Constructing the initial conditions based on peculiar velocities derived from the CosmicFlows-2 catalogue makes the predictions of the simulations completely independent from the distribution of the observed tracer population and following galaxy formation physics directly in the hydro-dynamical simulations allows to additionally base the comparison directly on stellar masses of galaxies or X-ray luminosity of clusters. We also used the 2668 $h^{-1}$Mpc large cosmological box from the {\it Magneticum} simulations to evaluate the frequency of finding such anomalies in random patches within simulations.}
  % results heading (mandatory)
   {We demonstrate that haloes and galaxies in our constrained simulation trace the local dark matter density field very differently. Thereby, this simulation reproduces the observed 50\% under-density of galaxy clusters and groups within the sphere of $\approx$100 Mpc when applying the same mass or X-ray luminosity limit used in the observed cluster sample (CLASSIX),
   which is consistent with a $\approx 1.5\sigma$ feature. At the same time, the simulation reproduces the observed over-density of massive galaxy clusters within the same sphere, which on its own also corresponds to a $\approx1.5\sigma$ feature. Interestingly, we find that only 44 out of 15635 random realizations (i.e. 0.28\%) are matching both anomalies,
   making the local Universe to be a $\approx3\sigma$ environment. We finally compared a mock galaxy catalogue with the observed distribution of galaxies in the local Universe, finding also a match to the observed factor of two over-density at $\sim 16$~Mpc as well as the observed 15\% under-density at $\sim$40~Mpc distance.}
  % conclusions heading (optional), leave it empty if necessary 
   {Constrained simulations of the local Universe which reproduce 
   the main features of the local density field open a new window for local field cosmology, where the imprint of the specific density field and the impact on the bias through the observational specific tracers can be investigated in detail. 
   }


   \maketitle
%
%-------------------------------------------------------------------

\section{Introduction}

The neighborhood in the immediate vicinity of the Milky Way (MW) is known as the “Local Group”. It is a binary system composed of two average-sized galaxies (the MW and Andromeda) that occupies a volume that is roughly $\sim$7 Mpc in diameter. At a distance of around 16 Mpc, the Virgo cluster comes into view as the main defining feature of our neighborhood on these scales. Beyond Virgo, a number of well-known and well-observed clusters like Centaurus, Fornax, Hydra, Norma, Perseus, and Coma dominate the local volume, among them a significant number of very massive clusters. 

Therefore, our local Universe, that is centered on us and extends over 150 $h^{-1}$Mpc, is not only a formidable site for detailed observations, but also appears as a very particular region of the Universe. Indeed, starting from a local void \citep{1987ang..book.....T}, bordered by the local sheet \citep{2008ApJ...676..184T}, there are also a large number of super-cluster structures identified within the local Universe, among them Perseus-Pisces, Centaurus, Coma and Hercules \citep[see recent work by][and references therein]{2021A&A...656A.144B}. Several of these most prominent structures in the local Universe form the so-called supergalactic plane, which was already recognized by \citet{1953AJ.....58...30D}. See also \citet{2000MNRAS.312..166L,2022MNRAS.511.5093P} for a summary of our current understanding of these structures as well as \citet{1986AcC....14....7F,1989woga.conf..431R} for historical reviews.
The impact of these structures is also recognized to form a global, pancake like structure out to a scale of $\sim$100~Mpc  \citep{1983IAUS..104..405E,1994MNRAS.269..301E,2021A&A...651A..15B} as well as large differences in the mean stellar density between the Northern and Southern hemispheres \citep{2018AN....339..615K} out to distances of $\sim$60~Mpc. It is often argued, that such particular structures show unusual over- \citep{2011MNRAS.412.2498M} or under- \citep{2020A&A...633A..19B} density when using luminous matter on different scales, with different conclusions when evaluating the underlying dark matter density field from them. As this could play a significant role in some of the current tensions in cosmology, for example the larger $H_0$ value locally inferred compared to CMB measurements \citep[see][for a most recent compilation of the values]{2021ApJ...919...16F}, it is important to understand how galaxies and galaxy clusters with different masses and properties trace the underlying dark matter density field. Also, the selection effect of different observations has to be understood in detail to answer such questions.

Trying to understand such features in the local Universe motivated various campaigns producing constrained simulations in the past. However, for a more detailed study one has to cover large enough volume, covering several hundreds of Mpc and initial conditions which are not directly based on the distribution of the traces to allow an independent comparison. In addition, one needs to include galaxy formation physics into such simulation to properly select objects by observables like stellar masses of galaxies or X-ray luminosity of galaxy clusters. Here we present simulations, which for the first time matches all the three criteria.

The paper is structured as follows: Section \ref{constrained_sim} gives an extended summary of previous constrained simulations followed by section \ref{simulations}, which describes the details of the used simulations, the included galaxy formation physics and details on the building of the initial conditions, followed by a qualitative comparison to observations in section \ref{qualitative_comparison}. In section \ref{haloes_ihn_slow}, we then present the predicted, spatial distribution of matter, halos and galaxies from the simulations and compare the peculiar features extensively with various observations in section \ref{comparison_with_observations}. In addition to the conclusions presented in section \ref{conclusions}, we also repeated some of the analysis on previous constrained simulations in the appendix \ref{comparison_sims} to demonstrate the potential inherent in our constrained simulations which does not depend on assuming a bias between tracer particles and dark matter but rather predicts this bias quite accurately.  

\begin{figure}
\centering
\includegraphics[width=0.99\hsize]{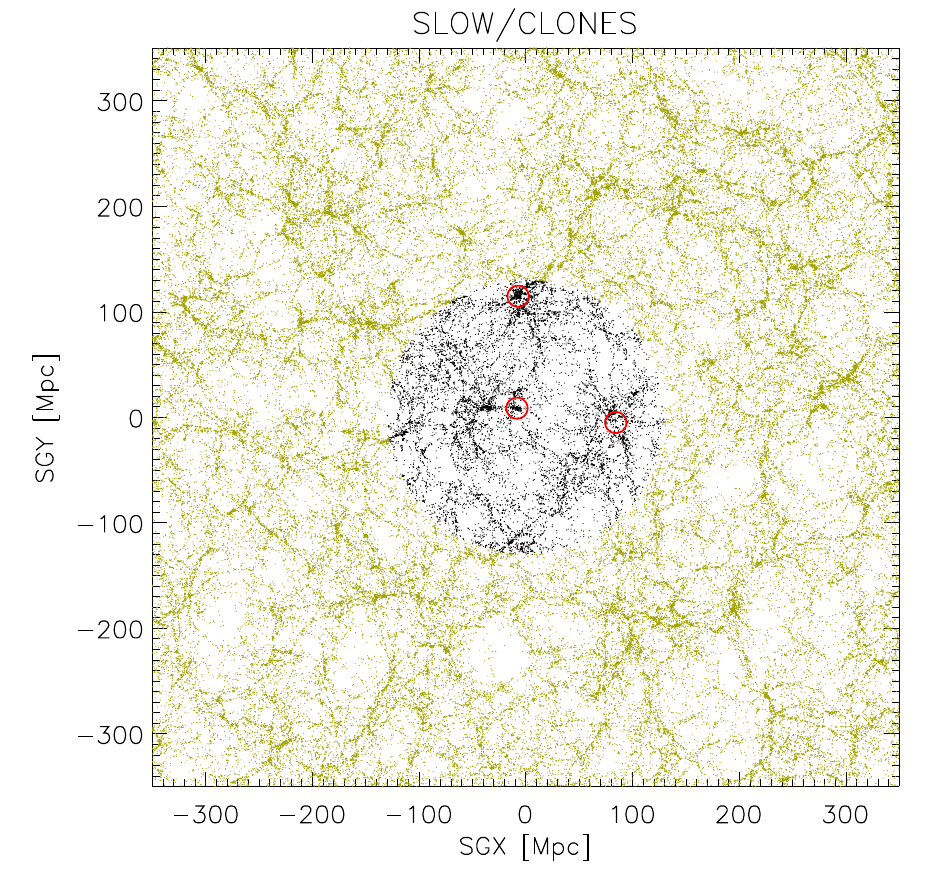}\\
\includegraphics[width=0.99\hsize]{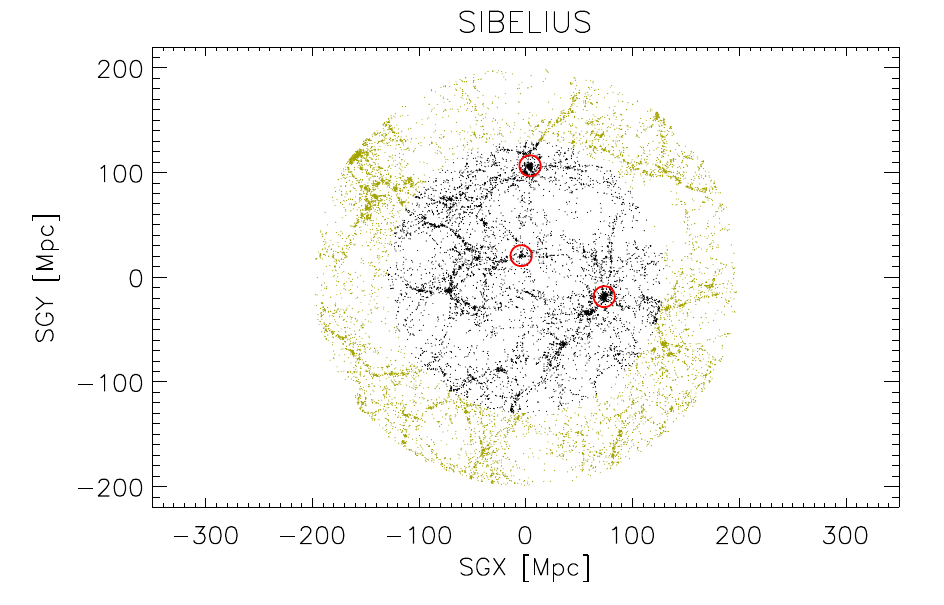}\\
\includegraphics[width=0.99\hsize]{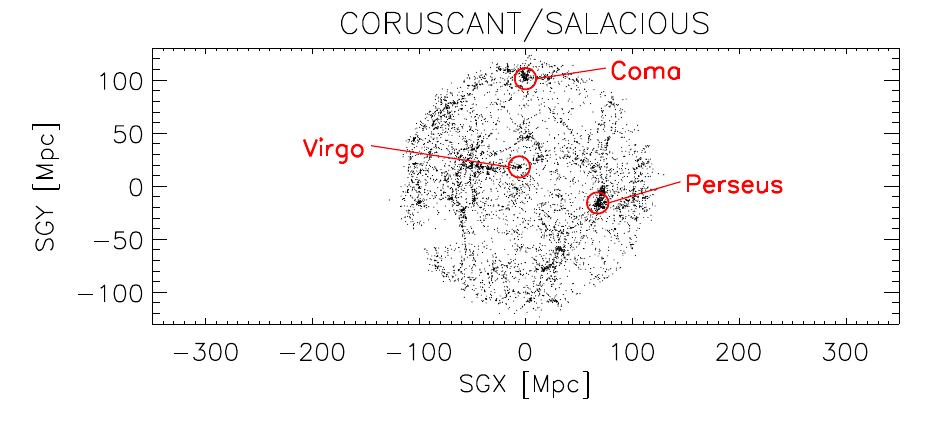}\\
\includegraphics[width=0.99\hsize]{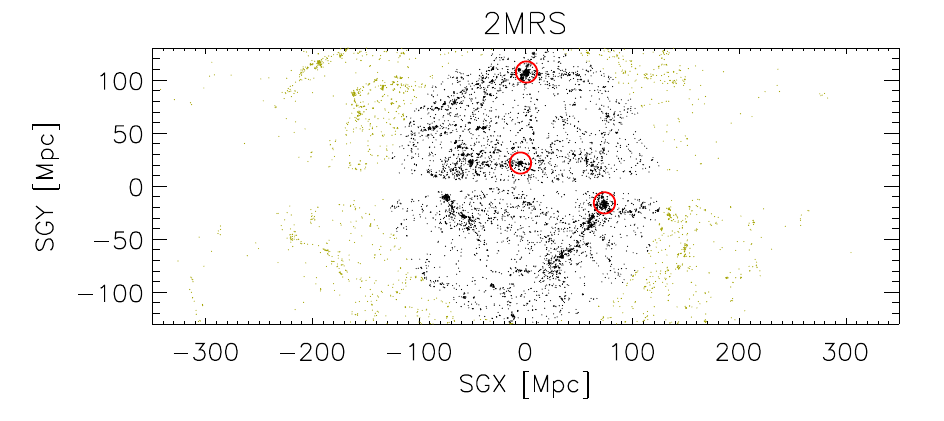}
\caption{Positions in super-galactic x/y coordinates of all galaxies in the simulations, compared to galaxies from the 2MRS catalogue. To make the simulations comparable, we restricted to galaxies with stellar masses above $10^{10}M_\odot$ within a 50 Mpc thick slice around the center of the simulation volume. We also used a lighter color for galaxies more distant than 130 Mpc. For the observation, we plotted only galaxies with $\mathrm{log}(L_K) > 10.25$ to match the mass cut. In addition, the positions of Coma, Virgo and Perseus are marked. The virial masses of the according halos in the simulations are listed in table \ref{tab:clustermass}. From top to bottom: SLOW, SIBELIUS, CORUSCANT  and 2MRS.}
\label{ComparisonSimsMaps}
\end{figure}

\section{Constrained Simulations}
\label{constrained_sim}

There are two approaches to study the problems  mentioned above with numerical simulations. One could run simulations of very large volumes with very high resolution and find the objects of interest in these simulations in similar environment as the observed ones. For example, in a statistical approach,  the  scatter in the $H_0$ value locally inferred can be studied in a box of (6 $h^{-1}$Gpc)$^3$  \citep{2014MNRAS.438.1805W} or Local Group candidates can be drawn from a large sample of isolated halo pairs  identified in a set of cosmological simulations \citep[APOSTLE,][]{2016MNRAS.457..844F}. An alternative approach is to use so-called constrained cosmological simulations. The goal of these simulations is to reproduce as well as possible the  positions, velocities, masses and internal properties of the objects of interest,   i.e. to reduce the cosmic variance in the region of interest as much as possible. Then such simulations can go well beyond the statistical approach and even trace back evolutionary paths of the studied objects. Naturally, constrained simulations are based on observations. There are two different approaches to obtain initial conditions of cosmological simulations based on the observed matter distribution at present. The first one goes backwards in time and is based on the Hoffman-Ribak algorithm \citep{1991ApJ...380L...5H}. We will use this algorithm throughout our paper. An alternative approach to study the local Universe with constrained simulations has been developed during the last decade, namely a Bayesian forward modelling 
(e.g.,\cite{2012MNRAS.427L..35K}, [ELUCID] \cite{2016ApJ...831..164W}, [SIBELIUS] \cite{2022MNRAS.509.1432S,2022MNRAS.512.5823M},  for a recent  comprehensive discussion of this Bayesian modeling see \cite{2019A&A...625A..64J}). The quality of the constrained simulations depends on the number and the quality of the constraints, see for example the discussion of the optimal sampling of velocity constraints for Wiener Filter reconstructions by \cite{2017MNRAS.468.1812S}. 

One of the first constrained simulations was based on data derived from redshift surveys, which estimate the local, total matter density from the distribution of galaxies using their observed redshift. Here, the main uncertainties are the assumed, constant relation between galaxy density and total density as well as the imposed uncertainty in the distances by presence of  peculiar line-of-sight velocity. Initial conditions for a 240 $h^{-1}$ Mpc box have been constructed, which contains a sphere of 160 $h^{-1}$ Mpc diameter sampled with higher resolution \citep{2002MNRAS.333..739M}. Simulations based on these initial conditions cover dark matter only simulations combined with semi analytic models where used to study the local galaxy population \citep{2002MNRAS.333..739M} and pure magneto-hydrodynamical simulations to study the propagation of cosmic rays in the local Universe \citep[CORUSCANT\footnote{\url{https://www.usm.lmu.de/~dolag/Simulations/\#CORUSCANT}},][]{2004JETPL..79..583D} as well as the imprint of the local super-cluster onto the CMB \citep{2005MNRAS.363...29D}. Subsequently, these initial conditions have been evolved with more advanced physics including radiative cooling, star-formation, stellar evolution and chemical enrichment to study density and temperature fluctuations in the local and prominent galaxy clusters \citep[SALACIOUS\footnote{\url{https://www.usm.lmu.de/~dolag/Simulations/\#SALACIOUS}},][]{2007ApJ...659..257K}. Including, additionally, AGN feedback from supermassive black holes, subsequent simulations based on these initial conditions have been used to contrast the observed Sunyaev-Zel'dovich effect (SZ) of the Coma and Virgo clusters as measured by the \textit{Planck} satellite \citep{2013A&A...554A.140P,2016A&A...596A.101P}. 

The other method to constrain initial conditions is based on peculiar velocities derived from direct distance measures. This has the advantage that tracing the velocity field is independent of the tracer population and that the velocity traces the potential caused by the large scale density field of all matter. However, the data sets of galaxies with measured distance indicators are much smaller and the intrinsically large uncertainties associated with these distance indicators have to be dealt with in a very careful way. Early attempts therefore have tried a hybrid approach, using velocity data combined with constraints coming from local galaxy clusters which allowed to span large enough volumes to cover the most important galaxy clusters like Virgo, Perseus and Coma. This led to the  CLUES\footnote{\url{www.clues-project.org}} \citep{2009AIPC.1178...64Y} dark matter only simulation of a 160 $h^{-1}$ Mpc box centered on the MW position to study the local galaxy population \citep{2003ApJ...596...19K} and a non-radiative, hydrodynamical simulation of the Virgo cluster within that box to study the properties of the intergalactic medium \citep{2002ApJ...571..563K}. Based on the same constraints, high resolution simulations of a 64 $h^{-1}$ Mpc box  \citep{2010arXiv1005.2687G}  have been performed and used for example to study the reionisation of the Local Group \citep{2018MNRAS.477..867D} while 
zoomed full hydrodynamical simulations within a few Mpc large, central region of the same box  have been used for many studies of the Local Group, including simulations following in detail the formation of  isolated dwarfs in the neighborhood of the Local Group \citep[see for example][]{2015MNRAS.450.4207B}.

\begin{table*}[ht!]
\centering
\begin{tabular}{c|ccc|ccccc|}
        & distance & 2MRS & PLANCK & LU2016 & \multicolumn{2}{c}{SLOW/CLONES} & CORUSCANT & SIBELIUS \\ 
        & $v_\mathrm{CMB}$ & $M_\mathrm{dyn}/1.12$  & $1.7\times{}M_{500c}^{SZ}$ & $M_\mathrm{vir}$ & $v_\mathrm{rad}$ & $M_\mathrm{vir}$ & $M_\mathrm{vir}$ & $1.2\times{}M_{200c}$ \\
Cluster & $[\mathrm{km}/\mathrm{s}]$ & $[M_\odot]$  & $[M_\odot]$ & $[M_\odot]$ & $[\mathrm{km}/\mathrm{s}]$ & $[M_\odot]$ & $[M_\odot]$ & $[M_\odot]$ \\ \hline
Coma    & 7264 & $1.4\times10^{15}$ & $1.2\times10^{15}$ &                    & 8316 & $1.8\times10^{15}$ & $7.6\times10^{14}$ & $1.5\times10^{15}$ \\
Perseus & 5155 & $1.5\times10^{15}$ & &             & 6343 & $1.0\times10^{15}$ & $1.3\times10^{15}$ & $3.3\times10^{15}$ \\
Virgo   & 1636 & $6.3\times10^{14}$ & $8.1\times10^{14}$ & $(6.5\pm1)\times10^{14}$ & 1434 & $9.8\times10^{14}$ & $5.5\times10^{14}$ & $4.3\times10^{14}$ \\
\end{tabular}
\vspace{2.ex}
\caption{Observational properties of Coma, Perseus and Virgo showing the radial velocity and two estimates of their virial mass $M_\mathrm{vir}$. The dynamical mass is taken from the Tully galaxy groups catalogue \citep{2015AJ....149..171T} and corrected down by 12\%, as needed to convert the zero velocity mass to virial mass \citep{2016MNRAS.460.2015S}. Alternatively we quote the masses inferred from scaling $M_{500c}$ for Coma from PLANCK data \citep{2013A&A...554A.140P}, following \citet{2021MNRAS.500.5056R} for converting the different masses or taking $M_\mathrm{vir}$ from the measured gas mass of Virgo as obtained from PLANCK data \citep{2016A&A...596A.101P}. We also show the virial masses obtained in the SLOW, CORUSCANT and SIBELIUS simulations, as well as the distribution from the 200 Virgo-like clusters from the LU2016 simulations \citep{2020MNRAS.496.5139S} and the radial velocity of the clusters in the SLOW simulation to be able to compare their radial distances. See also \citet{2018MNRAS.478.5199S} for the variance of the cluster properties in different CLONES realizations. Note that we converted $M_{200c}$ as given for the SIBELIUS simulation to $M_\mathrm{vir}$ following \citet{2021MNRAS.500.5056R} and using the same cosmology as that used in the SIBELIUS simulation. A more detailed comparison of cluster properties from the SLOW simulation with observations will be presented in Hern\'andez-Mart\'inez et al. (in prep).} 
\label{tab:clustermass}
\end{table*}

The growing data of direct distances from the CosmicFlow project \citep[e.g.][]{2012ApJ...744...43C,2013AJ....146...86T,2016AJ....152...50T,2022arXiv220911238T} has driven new approaches within the CLUES project since then. In a series of papers, techniques have been developed to deal with the increasing number of constraints and to handle the biases inherent to velocity data \citep{2013MNRAS.430..888D,2013MNRAS.430..902D,2013MNRAS.430..912D,2014MNRAS.437.3586S,2016MNRAS.455.2078S,2015MNRAS.450.2644S}. Based on this approach, initial conditions for a 64 $h^{-1}$ Mpc box have been constructed. They have been used within the CosmicDawn project to run a fully coupled radiation-hydrodynamics simulation of cosmic reionisation and galaxy formation until redshift $z=6$ \citep{2020MNRAS.496.4087O} as well as to run a dark matter only simulation with $4096^3$ particles down to redshift $z=0$ within the MultiDark project. In this simulation, Milky Way and Andromeda galaxies can be identified so that their reionisation history can be inferred from the radiation-hydrodynamics simulation \citep{2022MNRAS.515.2970S}. Moreover, initial conditions based on the same approach have been used to simulate a 100 $h^{-1}$ Mpc box, where a high resolution region of about 10 $h^{-1}$ Mpc is placed in the central region around the local group, including detailed galaxy formation physics \citep[HESTIA,][]{2020MNRAS.498.2968L}. First dark matter only simulations based on a larger  500 $h^{-1}$ Mpc box already allowed to reproduce the Virgo cluster and its formation history \citep[LU2016,][]{2016MNRAS.460.2015S}. Consecutive improvement of dark matter simulation of this volume allowed to reproduce more and more prominent galaxy clusters like Perseus and Coma \citep[CLONES, Constrained LOcal and Nesting Environment Simulations,][]{2018MNRAS.478.5199S}, to study the formation history of clusters \citep{2018A&A...614A.102O,2020MNRAS.496.5139S}, and to study the large-scale cosmic web in which clusters like Coma are embedded by comparing observations \citep{2020A&A...634A..30M} and simulations \citep{2023arXiv230603124M}. Subsequent placing high resolution regions around such prominent clusters allows direct comparison of galaxy properties within the Virgo cluster from hydrodynamical simulations with full galaxy formation physics with their observed counterparts \citep{2021MNRAS.504.2998S}. For the first time, the CLONE (Constrained LOcal and Nesting Environment) simulation results undoubtedly show that the constrained formation history of the Virgo cluster significantly differs from averaged clusters of the same mass \citep{2021MNRAS.504.2998S}. Moreover, the simulations indicate phase-space positions of recently in-falling galaxy groups. It also matches the specific amplitude and shapes of the appearing velocity waves on the lines-of-sight towards various, well known galaxy clusters \citep{2023arXiv230101305S}.

The SLOW (Simulating the Local Web) simulation\footnote{\url{https://www.usm.lmu.de/~dolag/Simulations/\#SLOW}} is a 500 $h^{-1}$Mpc box, using the realization number 8 of CLONES, as described in Sect. \ref{slow} and assumes a \textit{Planck} like cosmology \citep{2014A&A...571A..16P}, with a Hubble constant $H_0=67.77 \mathrm{km/s/Mpc}$, a total matter density of $\Omega_\mathrm{Matter}=0.307115$, a cosmological constant of $\Omega_\Lambda=0.692885$ and a baryon fraction corresponding to $\Omega_\mathrm{baryon}=0.0480217$. It follows the evolution of dark and baryonic matter within a (500 $h^{-1}$Mpc)$^3$ simulation volume centred on the position of the MW and that stands for our cosmic neighborhood. The initial conditions for such simulations are obtained with sophisticated algorithms (see section \ref{slow}) that take into account the position and motion estimates of thousands of galaxies within our local volume. These local measurements allow us to constrain the initial conditions that, in turn, lead to the observed local large scale structure, when motions since early times until today are followed according to the gravity laws. In addition, the baryonic matter is treated via hydro-dynamics together with various, state-of-the-art sub-grid models (see section \ref{galform_model}). There are variants which follow the evolution of the magnetic field and cosmic rays, others are following the formation of the stellar population as well as black holes (BHs) and associated active galactic nuclei (AGN) physics. Here we are following the prescriptions as used for the {\it Magneticum} simulations (see \cite{2014MNRAS.442.2304H,2016MNRAS.463.1797D}). This first paper in a series mainly introduces the general properties, based on the simulation following galaxy formation processes and therefore a realistic galaxy population is present in the simulations as well as galaxy clusters with proper intracluster medium (ICM) properties. Therefore, galaxies together with the population of galaxy clusters can be directly compared by the same means to the observed counterparts. This also allows us to study the effects of the local environment on observational properties and their cosmological impact.

However, comparing clusters from constrained simulations with their observational counterparts need a cross-identification, which itself is subject to evaluating differences in observed and simulated positions and masses. Such differences in positions have different origins. Simulations performing density reconstruction based on redshift survey typically need to apply a relatively large smoothing (e.g. 5-10 Mpc as in the case of CORUSCANT) to the reconstructed density field, while the bulk velocity with which the halo is moving over cosmic time is largely unconstrained and leads to additional displacement of the halo in the simulation. In simulations based on peculiar velocities, uncertainties from distance moduli are propagating to radial velocities and, when applying Poisson equation, propagate further into uncertainties on the reconstructed density / total velocity field. Therefore uncertainties in the observed distance (which could be as large as several tenth of Mpc for distant clusters) are coupled directly with displacements of the three dimensional positions within the evolved simulation.

To give a better impression on the differences in some of the constrained simulations, we show the predicted distribution of galaxies from the three simulations SLOW (upper panel), SIBELIUS and SALACIOUS (middle panels) compared to the galaxies from the 2MRS catalog (lower panel) in figure \ref{ComparisonSimsMaps}. In addition, the positions of Coma, Perseus and Virgo within the simulations and observations are shown. In general, the environment of these clusters are similar in the different simulations (like the large scale structure leaving Perseus to the lower left, or the horizontal structures leaving Virgo and Coma. However, the details in these structures are different as well as there are striking differences between the massive clusters and their immediate environment. In SLOW/CLONES for example, the immediate outskirts of Perseus indicate a quite fossil, relax system with a mass which is very close to that expected from observations, where Perseus is characterized by a strong cool-core feature which is associated with overall relaxed systems \citep[see][]{1981ApJ...248...47F,1993MNRAS.264L..25B}. In contrast, within the CORUSCANT/SALACIOUS simulation the Perseus structure appears as a twin system with an even more massive companion, while in the SIBELIUS simulation Perseus appears to be three times too massive. The obtained virial masses of these clusters in the different simulations and their positions are listed in table \ref{tab:clustermass}, where we also listed the observational findings. For the masses, we converted all values to the same mass definition using the scaling relations given in \citet{2021MNRAS.500.5056R} and \citep{2016A&A...596A.101P}. 

\begin{figure*}[ht!]
\centering
\includegraphics[width=0.99\hsize]{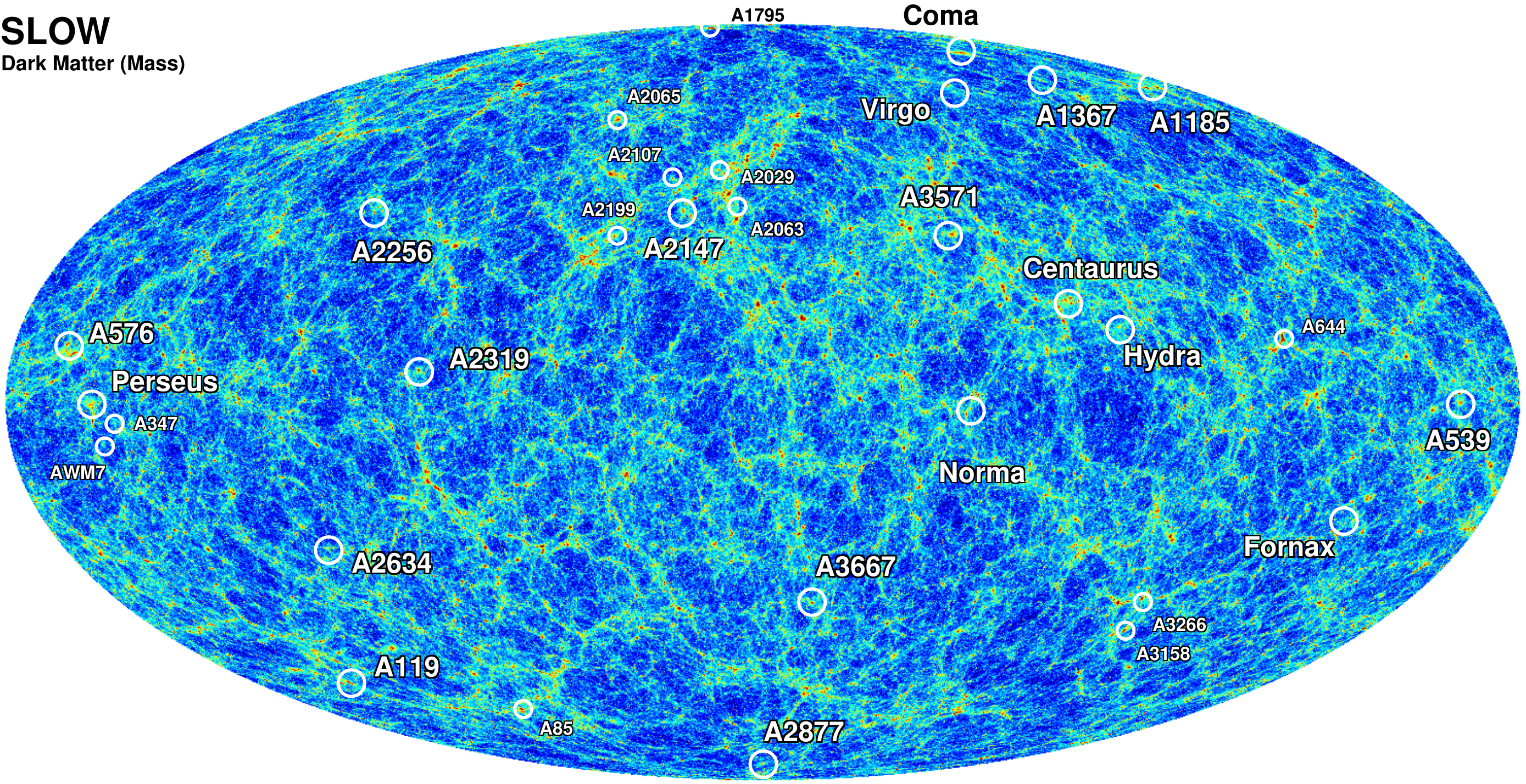}
\caption{Shown is a full sky map of the projected dark matter distribution from the SLOW simulation up to a distance of 350 Mpc. To distinguish the constrained part of the local Universe simulation, we labeled some of the cross-identified clusters in the Local Universe.}
\label{FigFullSky_dm}
\end{figure*}

\begin{figure*}[htp!]
\centering
\includegraphics[width=0.99\hsize]{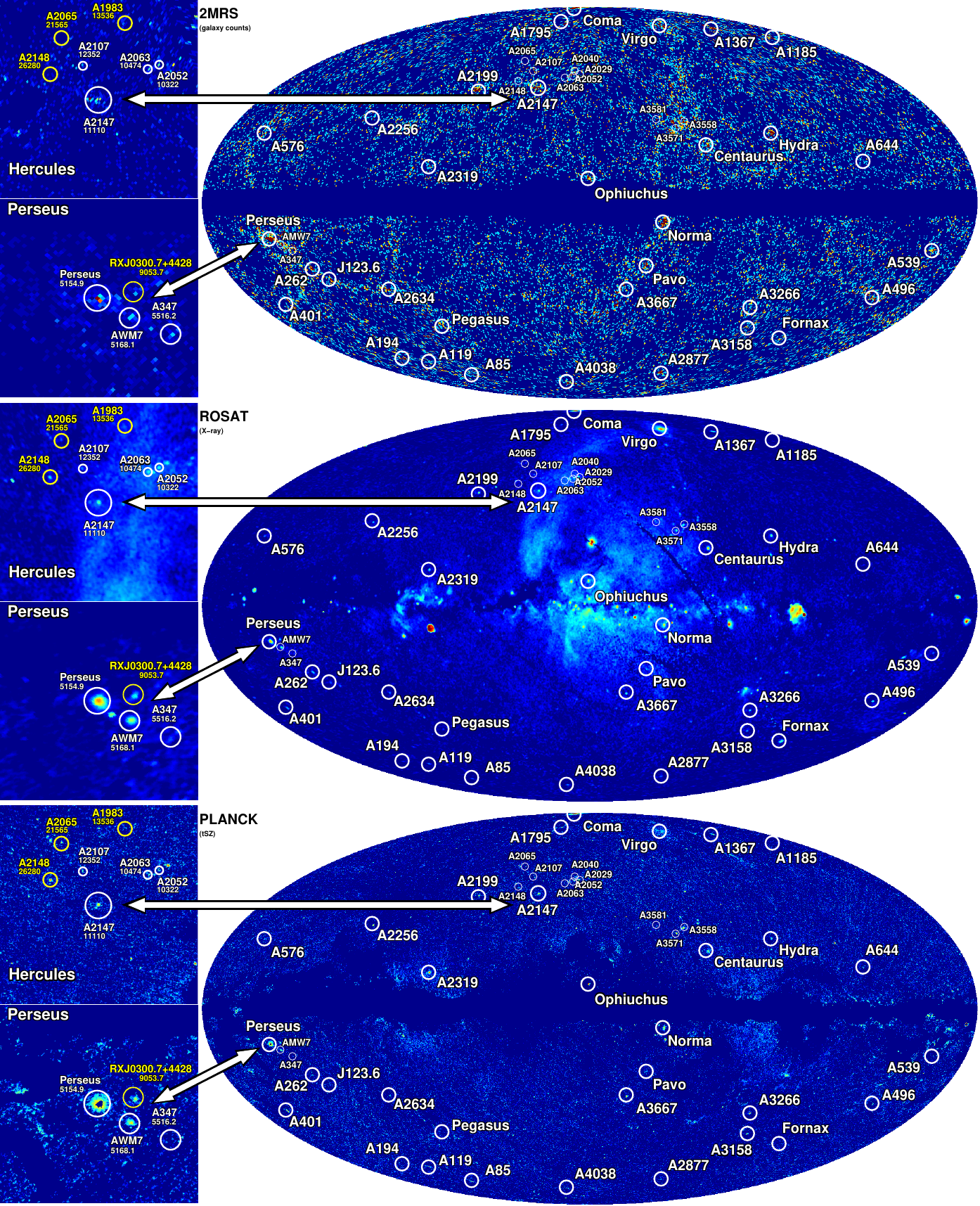}
\caption{The different panels show observations in different wavebands. From top to down: distribution of galaxies in the 2MRS catalog, X-ray surface brightness in the ROSAT's 1.5 keV band (R6+R7=0.76-2.04~keV) and Compton-Y map based on \textit{Planck} data \citep[produced with the MILCA algorithm][]{2016A&A...594A..22P} with the CO mask by \cite{2016A&A...592A..48K}. The inlays in addition show a zoom on the Perseus and Hercules regions. The labels in the inlays are giving the Name together with radial velocity (in km/s) as redshift distance indicator. The yellow labels are indicating clusters which are outside the slice used in the simulations shown in figure \ref{FigFullSky_sim}.}
\label{FigFullSky_obs}
\end{figure*}

\begin{figure*}[htp!]
\centering
\includegraphics[width=0.99\hsize]{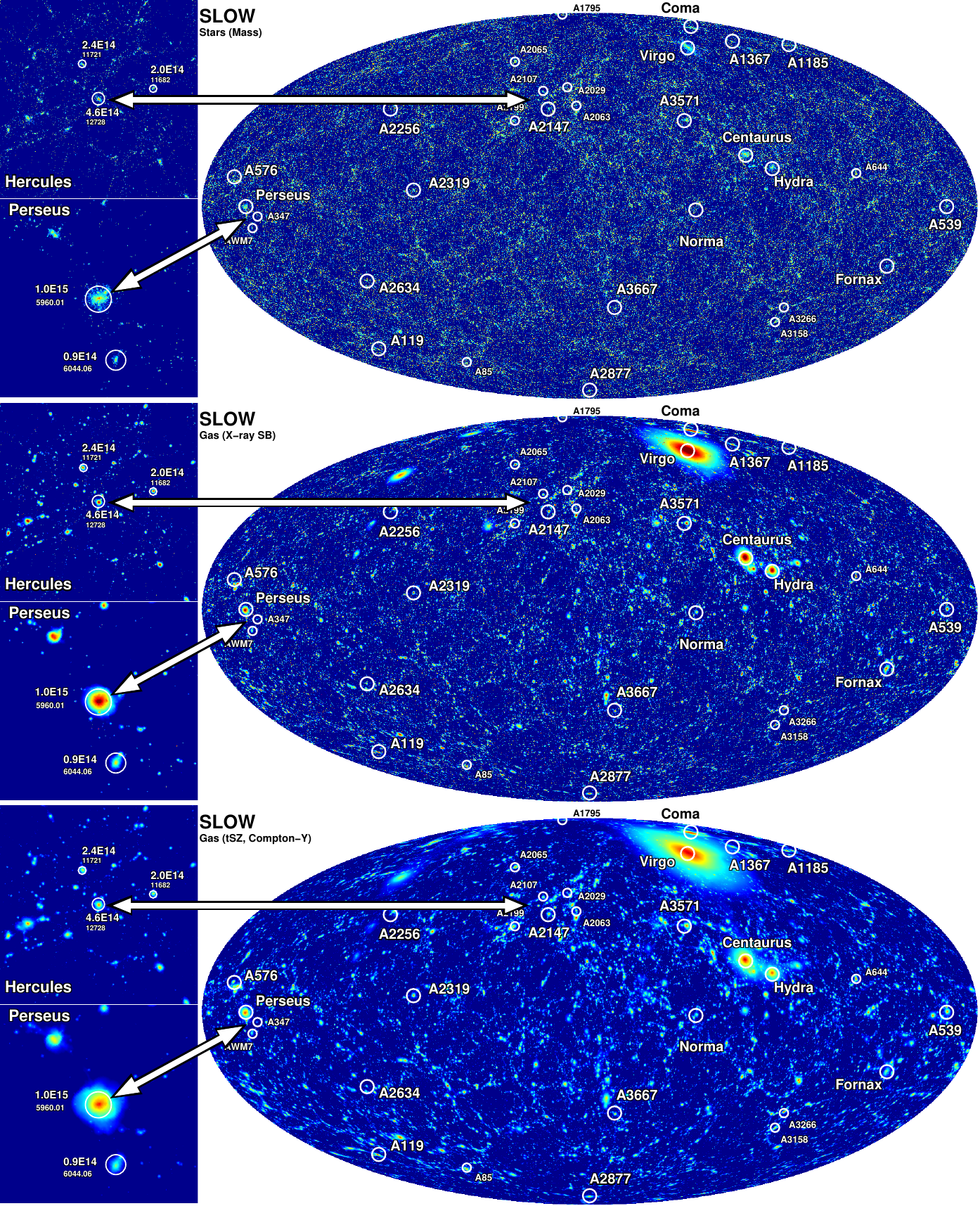}
\caption{The different panels show the SLOW simulation in different wavebands. From top to down: distribution of stellar mass in SLOW, X-ray surface brightness and Compton-$Y$. We used here a much larger dynamical ranges than in the observations shown on figure \ref{FigFullSky_obs} for the color scaling to emphasize the large angular imprint on the very local galaxy clusters. Note also that we here always show the full simulation volume, out to a distance of 350 Mpc. The inlays in addition show a zoom on the Perseus and Hercules regions. Here we used a much more narrow range around Perseus and A2147, e.g. a 10 respectively 40 Mpc thick region to emphasize the local structures. The labels in the inlays are giving the virial mass (in $M_\odot$) of the haloes, together with radial velocity (in km/s) as redshift distance indicator.}
\label{FigFullSky_sim}
\end{figure*}

%--------------------------------------------------------------------
\section{Simulations}
\label{simulations}

\subsection{Magneticum simulations}
\label{magneticum}

The {\it Magneticum}\footnote{\url{www.magneticum.org}} Simulations follow for the first time the evolution of up to 10$^{11}$ particles in a series of hydro-dynamical simulations of cosmological volumes ranging in size from (68 Mpc)$^3$ to almost (4 Gpc)$^3$. To evaluate the significance of features in the local Universe, we use the largest simulation volume, {\it Box0/mr} of the {\it Magneticum} simulation set as reference for the general presence of such features within the $\Lambda$CDM cosmological model. This simulation covers a box of 2.688$h^{-1}$GPc in size, resolved with a total of $2\times4536^3$ dark matter and gas resolution elements, having a mass resolution of $1.3\times10^{10}h^{-1}M_\odot$ and $2.6\times10^9h^{-1}M_\odot$ respectively and featuring full galaxy formation physics. It is therefore well equipped to reflect galaxy cluster and group properties. Having $\approx 150$ times the volume of the 500$h^{-1}$Mpc box, {\it Box0/mr} of the {\it Magneticum} simulations allow us to compare a large number of random patches with the constrained part of the SLOW simulation with a very high statistical significance. The cosmology adopted for these simulations is slightly different than for SLOW, as it follows the WMAP7 from \citet{2011ApJS..192...18K}, with Hubble constant $H_0=70.4 \mathrm{km/s/Mpc}$, a total matter density of $\Omega_\mathrm{Matter}=0.272$, a cosmological constant of $\Omega_\Lambda=0.728$ and a baryon fraction corresponding to $\Omega_\mathrm{baryon}=0.0459$, as well as an overall normalisation of the power spectrum of $\sigma_8=0.809$ and a slope of the primordial fluctuation spectra of $n=0.963$. However, these small differences do not play any significant role for the comparisons presented in this study. 

\subsection{The sub-grid model used}
\label{galform_model}

Both, the SLOW as well as the {\it Magneticum} simulations are using an updated formulation of SPH \citep{2016MNRAS.455.2110B} with modern, high order Kernels \citep{2012MNRAS.425.1068D} and include the treatment of the relevant models to describe the physical processes needed for galaxy formation like cooling, star formation and winds. This is based on the multi-phase model \citep{Springel2003}, but extended to follow in detail the stellar population and chemical enrichment by SN-Ia, SN-II, AGB  \citep{2003MNRAS.342.1025T,2007MNRAS.382.1050T}, uses metal depending cooling tables from \citet{2009MNRAS.399..574W} and a galactic wind velocity of $\rm 350 \ km/s$ for the kinetic feedback. Another important aspect to follow are the evolution of super massive black holes and the associated AGN feedback, where we follow \citep{2005MNRAS.361..776S} with various improvements \citep{2010MNRAS.401.1670F,2014MNRAS.442.2304H} for the treatment of the black hole sink particles and the different feedback modes. Importantly for the treatment of the ICM they include isotropic thermal conduction of 1/20 of standard Spitzer value \citep{2004ApJ...606L..97D} and a low viscosity scheme to track turbulence \citep{2005MNRAS.364..753D,2016MNRAS.455.2110B}.

 It has been intensively demonstrated, that this sub-grid model leads to galaxy and ICM properties in galaxy clusters, which largely follow the observational trends and properties. Thereby, the {\it Magneticum} simulations have been compared to Sunyaev-Zel’dovich (SZ) data from \textit{Planck} \citep{2013A&A...550A.131P} and SPT \citep{2014ApJ...794...67M}. It has been demonstrated to reproduce the observable X-ray luminosity-relation \citep{2013MNRAS.428.1395B}, the chemical composition \citep{2017Galax...5...35D,2018SSRv..214..123B} of the ICM and the high concentration observed in fossil groups \citep{2019MNRAS.486.4001R}. On larger scales, the \textit{Magneticum} simulations demonstrated to reproduce the observed SZ power spectrum \cite{2016MNRAS.463.1797D} as well as the observed thermal history of the Universe \citep{2021PhRvD.104h3538Y} and the gas properties in between galaxy clusters \citep{Biffi22}. On galaxy scales, the simulations lead to an overall successful reproduction of the basic galaxy properties, like the stellar mass function at low \citep{2017ARA&A..55...59N} and high 
 \citep{2023MNRAS.518.5953L,2022arXiv220801053R} redshifts, the environmental impact of galaxy clusters on galaxy properties \citep{2019MNRAS.488.5370L,2023MNRAS.518.5953L}, the azimuthal distribution of matter around clusters compared with findings in SDSS \citep{2020A&A...635A.195G}  
 and the appearance of post-starburst galaxies \citep{2021MNRAS.506.4516L} as well as the associated AGN population at various redshifts \citep{2014MNRAS.442.2304H,2016MNRAS.458.1013S,2018MNRAS.481.2213B}. 
 
\subsection{SLOW}
\label{slow}
The large-scale structure of the Universe is effectively described by the (peculiar) velocity and real space distribution of observable galaxies. A Wiener Filter (WF) algorithm\footnote{Linear minimum variance estimator, in abridged form WF \citep{1995ApJ...449..446Z,1999ApJ...520..413Z}.} is needed to reconstruct the true underlying cosmography from the noisy and incomplete galaxy data reaped from surveys. The first attempt to construct constrained realizations of Gaussian random fields subject to linear constraints was made by \cite{1991ApJ...380L...5H}. In the intervening two decades the technique as well as the input constraints have been refined – peculiar velocity, in fact distance modulus, surveys such as CosmicFlows-2 \citep[CF2,][] {2013AJ....146...86T} are particularly useful since the cosmic velocity field is directly related to the matter density field in the linear regime. Wiener Filter reconstructions of CF2 have already been carried out successfully estimating the density field within $\sim$100 Mpc \citep[e.g. Laniakea,][]{2014Natur.513...71T}. \cite{2018MNRAS.478.5199S}  describes in details the steps of the method to build the constrained initial conditions and introduces those used in this paper. The main steps are summarized hereafter:
 \begin{enumerate}
 \item Before deriving the peculiar velocities, grouping \citep{2018MNRAS.476.4362S} of the distance modulus catalog to remove non-linear virial motions that would affect the linear reconstruction obtained with the linear method (e.g. \cite{2017MNRAS.468.1812S,2017MNRAS.469.2859S}). Typically, when several distance moduli are available for several galaxies within the same galaxy cluster, they are replaced by the distance modulus of the cluster.
 \item Minimizing the biases \citep{2015MNRAS.450.2644S} inherent to any observational radial peculiar velocity catalog (e.g. Malmquist biases and lognormal error). An iterative algorithm permits retrieving the Gaussian radial peculiar velocity distribution, expected from the theory, from a flat distribution with large tails. Additionally, uncertainties are derived for these new peculiar velocities to filter the smoothing effect with the distance (dilution of the information with the distance) of the subsequent algorithms \citep{2016MNRAS.455.2078S,2018MNRAS.478.5199S}.
 \item Reconstructing the cosmic displacement field with the WF technique applied to the peculiar velocity constraints.
 \item Accounting for the cosmic displacement by relocating constraints to the positions of their progenitors using the Reverse Zel’dovich Approximation and the reconstructed cosmic displacement field \citep{2013MNRAS.430..888D,2013MNRAS.430..902D,2013MNRAS.430..912D} and replacing noisy radial peculiar velocities by their WF 3D reconstructions \citep{2014MNRAS.437.3586S}.
 \item Producing density fields constrained by the modified observational peculiar velocities combined with a random realization to restore statistically the missing structures using the Constrained Realization technique \citep[CR,][]{1991ApJ...380L...5H,1992ApJ...384..448H}.
 \item Rescaling the density fields to build constrained initial conditions\footnote{\textsc{GINNUNGAGAP}: \url{https://github.com/ginnungagapgroup/ginnungagap}}, where increasing the resolution implies adding random small scale features.
\end{enumerate}

Here, as well as for the actual simulations, we assume the standard, $\Lambda$CDM cosmological model with the Hubble constant $H_0=67.77 \mathrm{km/s/Mpc}$, a total matter density of $\Omega_\mathrm{Matter}=0.307115$, a cosmological constant of $\Omega_\Lambda=0.692885$ and a baryon fraction corresponding to $\Omega_\mathrm{baryon}=0.0480217$, as well as an overall normalisation of the power spectrum of $\sigma_8=0.829$ and a slope of the primordial fluctuation spectra of $n=0.961$ \citep{2014A&A...571A..16P}.

\subsection{SLOW set of simulations}
\label{slow_sims}
 
To create the initial conditions, SLOW use the CLONE-500~Mpc/h-512$^3$grid, realization number 8, to which different small scale features for different resolutions are added with \textsc{Ginnungagap}. Simulations are then performed with different levels of additional physics: 
\begin{itemize}
\item Dark matter only simulations were performed using $768^3$, $1576^3$, $3072^3$ and $6144^3$ particles.
\item Magneto-hydrodynamical, non radiative simulations but following also a cosmic ray component \citep{2023MNRAS.519..548B} using $2\times3072^3$ gas and dark matter particles.
\item Hydrodynamical, full galaxy formation physics simulation using $2\times768^3$, $2\times1576^3$ and $2\times3072^3$ gas and dark matter particles (last one only down to $z=2$).
\end{itemize}

Halos are identified using \textsc{SubFind} \citep{2001MNRAS.328..726S,do09} which detects halos based on the standard Friends-of-Friends algorithm \citep{1985ApJ...292..371D} and subhalos as self-bound regions around local density peaks within the main halos. The centre of halos and subhalos are defined as the position of the particle with the (local) minimum of the gravitational potential. The virial mass, $M_\mathrm{vir}$ is defined through the spherical-overdensity around a halo as predicted by the generalised spherical top-hat collapse model \citep{1996MNRAS.282..263E} with an over-density for the chosen cosmology following \citet{1998ApJ...495...80B}. For subhaloes, individual properties are computed based on the particles which are associated to the individual subhalos. To compute the K-band magnitudes in our simulation which include galaxy formation physics, we used the K-band stellar mass to light ratio as obtained from SDSS \citep{2003ApJS..149..289B}
\begin{equation}
\mathrm{log}_{10}(M/L_K) = -0.42 + 0.033\mathrm{log}_{10}(Mh^2/M_\odot),
\end{equation}
while in the case of dark matter only simulations, we are using the corresponding Tully-Fisher relation based on the maximum circular velocity computed within the subhaloes.

\begin{figure*}[t!]
\centering
\includegraphics[width=0.99\hsize]{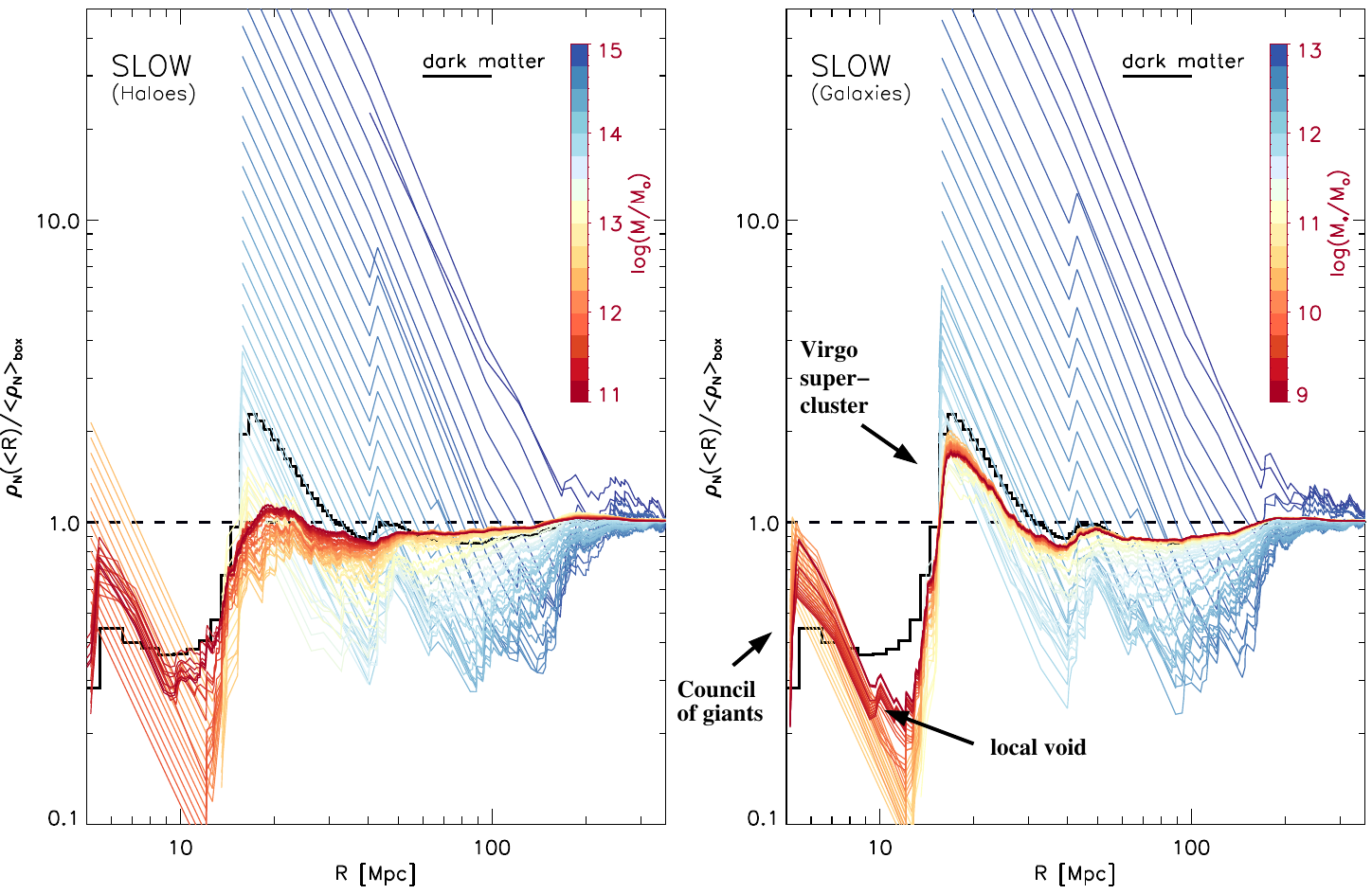}
\caption{Cumulative, relative number density of halos (left) and number density of galaxies (right) as function of distance from the MW position as obtained from the SLOW simulation. The different colors are lines obtained for different lower mass cuts, as indicated in the color bars: virial masses of the halos (left) and stellar masses of the galaxies (right). The black line in both panels marks the same when using directly dark matter distribution. 
}
\label{DensityHaloes}
\end{figure*}

\section{A qualitative comparison with observations}
\label{qualitative_comparison}

Within the SLOW simulations, more than 40 local clusters have been cross identified with their observational counterpart, showing good agreement in global properties (like total mass) as well as profiles of the ICM properties (like pressure and temperature), see Hern\'andez-Mart\'inez et al. (in prep) for details. All the results presented in this paper are based on the hydrodynamical, full galaxy formation simulation with $2\times1576^3$ particles. In Figure~\ref{FigFullSky_dm} we show the distribution of the dark matter within the SLOW simulation as a full sky map up to a distance of 350 Mpc, where the color coding is according to the total matter content in each pixel. This shows how the constrained region of the local Universe (where we labeled a subset of the cross identified clusters) is embedded in the larger cosmological structures. In Figure~\ref{FigFullSky_obs} we show a collection of observations. First, we show the distribution of galaxies in the 2MRS catalog. Note that here the observations are limited to more local galaxies as the 2MRS gets very sparse at distances beyond 100 Mpc. On the other hand, X-ray and Compton-$Y$ map contain significant foreground and other, non-cluster related emission. Here the selection of the 1.5 keV (R6+R7=0.76-2.04 keV) band for the X-ray surface brightness map, based on the data of the ROSAT All-Sky Survey \citep[RASS,][]{1997ApJ...485..125S}, maximises the visibility of the cluster signal. In the Compton-$Y$ map, the exclusion of regions with significant CO emission \citep[according to the mask by ][]{2016A&A...592A..48K}  from the \textit{Planck} data \citep[produced with the MILCA algorithm][]{2016A&A...594A..22P} is reducing the non-cluster related foreground. Therefore one has to keep in mind that both observational tracers contain objects which are outside the distance range of the simulation counterparts (some prominent ones are marked with yellow labels), and even some much more distant objects, even outside the simulation volume, as well as residuals of local, galactic foreground features and are limited on the faint end by the observational noise. This can be compared with the according counterparts as derived from the simulation, shown in figure \ref{FigFullSky_sim}. Here we show the distribution of the stellar component (upper panel), the X-ray surface brightness (middle panel) and the Compton-$Y$ map (lower panel) as obtained from the SLOW simulation. The maps are created from the simulations using SMAC \citep{2005MNRAS.363...29D}, where for the X-ray map the emissivity for each SPH particle is computed following \citet{1996MNRAS.283..431B}. Here we can use a much larger dynamical ranges in the color scaling to emphasize the large angular imprint on the very local galaxy clusters. Also here we always used the full simulation volume out to the distance of 350 Mpc for producing the counterpart sky maps from the simulations. In addition, we show two special regions (Perseus and Hercules) in more detail in the inlays. Therefore we always center on the cross identified halo of the prominent cluster (e.g. Perseus and A2147) and used a much more narrow range around them, e.g. a 10 respectively 40 Mpc, to emphasize the local structures. The virial mass (in solar masses) are given as labels for the cross identified halos, while in addition the radial velocity (in km/s) is given as a distance indicator. A more detailed comparison of the individual cluster properties across multiple wavebands will be released in a series of papers. In table \ref{tab:clustermass} we give the virial masses for Coma, Virgo and Perseus as a reference. As can be seen when comparing the full sky maps from the observations and the simulations, uncertainties in the constrains which went into the construction of the initial condition lead to a noticeable shift in the positions \citep[see also][]{2018MNRAS.478.5199S}. The largest contribution to this positional discrepancies are the still relatively large uncertainties in the observed distance modules. This gets even more evident when comparing the inlays. For example in the Hercules region, A2147 has a clear match and also A2107 has a corresponding halo. There is also a halo resembling the A2063/A2052 complex, but this is only a single halo in the simulation and shows also a significant shift (e.g. several degrees on sky). A similar situation arises for the Perseus region. Perseus itself has a quite well matching counterpart, however, a possible counterpart to AWM7 is already significantly displaced. Nevertheless, this comparison across multiple wavebands clearly reveals that appearance of clusters can look significantly different across the different wavebands and demonstrate the need of full, hydro dynamical simulations for such comparison, as the apparent significance of structures often largely differs when comparing galaxy and ICM properties. Still, overall clear similarities in the large scale structure and the appearance of galaxy clusters are visible across the different wavebands when comparing the SLOW simulations with the observations, confirming \citet{2018MNRAS.478.5199S} and \citet{2023arXiv230101305S} assertions.
 
\section{Mean density for Halos in SLOW}
\label{haloes_ihn_slow}

There are various density anomalies reported in the Local Universe, ranging from the Council of Giants \citep{2014MNRAS.440..405M}, the local void \citep{1987ang..book.....T}, close structures, a local deficit of galaxy clusters \citep{2020A&A...633A..19B} and an overabundance of very massive galaxy clusters like Norma, Perseus, Coma, Ophiuchus and A2199 or A119. As SLOW is a constrained simulation, based on peculiar velocity observations, we can investigate how these structures are present in the predicted density field. Given the large volume of (500 $h^{-1}$Mpc)$^3$ covered by the simulation, we can investigate the full range of observed anomalies, given the resolution one can nowadays reach for such volumes. Having a full galaxy formation physics run allows also distinguishing between the dark matter density field, halos and galaxies, including observational properties like stellar mass for galaxies and properties of the intracluster medium (ICM) within galaxy clusters like their temperature, X-ray luminosity and more.

Figure~\ref{DensityHaloes} shows the cumulative number density within spheres with growing radius centered on the position of the MW. Here the left panel is using the distribution of halos with different lower limits in their virial mass as indicated by the legend, while the right panel uses the galaxies (e.g. sub-halos) with different lower limits in stellar mass. In both cases the spacing of the lines corresponds to a change in the mass threshold of $\Delta\mathrm{log}_{10}(M)=0.05$. The solid black line is obtained directly from the dark matter particles within the simulation.

Several features are immediately visible. Starting with a quite empty region within the very close vicinity around the MW position\footnote{Here we defined the position of the MW in this realization, so that the Virgo cluster is exactly at the correct position.}, there are several relatively massive galaxies at a distance of 5 Mpc, which build the equivalent of the so called Council of Giants, which are observed at $\approx$3.75 Mpc \citep{2014MNRAS.440..405M}. After that, the local void is clearly visible and filling the space till the Virgo galaxy cluster at $\approx$16 Mpc and the associated super cluster (the Virgo super cluster) comes into place. Then, between $\approx$ 30 and 140 Mpc, a clear under density is present in all tracers, except the very massive galaxies and massive galaxy clusters, which is in agreement with our findings based on the halo mass functions as presented in \citep{2016MNRAS.455.2078S}. he high start of the blue, upper lines reflect the fact that we have a very massive cluster like Virgo very close. Note that a typical, mean separation of clusters with a virial mass of $10^{15}M_\odot$ is $\approx 180$ Mpc. These lines stay high, as we approach other very massive clusters (among them Perseus and Coma) well before the distance reflecting the mean density of such systems. The main obstacle in relating the density of tracers to the underlying dark matter distribution depends on the physics of gravitational clustering and the more complex physics of galaxy formation and is often referred generically as "bias" \citep[see for example][and references therein]{2004ApJ...601....1W}. It is interesting to note that the bias between the tracer population and the dark matter for the galaxy population in the normal mass-range shows the expected, regular behaviour with only mild dependency on actual mass. In contrast, the bias using halos shows a strong mass dependency and also large fluctuations in relative amplitude when comparing halos to the dark matter distribution. It is equally interesting to note that the factor of 2 over-density associated with the Virgo supercluster structure as seen in dark matter shows only up in galaxies and only if galaxies down to stellar masses of at least $10^{11}\mathrm{M}_\odot$ are used and does not show up in the halo number density distribution at all. This means that in this case, the halo of the Virgo clusters seems generally quite isolated and therefore the halo itself traces the general over-density, but not the associated structures in the environment. At the very high mass end, galaxies and halos align, which just reflects that the central, very massive galaxies (BCGs) are strictly related to massive galaxy clusters and their halo, marking the point  where galaxies of a certain mass can no longer be formed by internal processes but mainly grow by (mostly dry) mergers.

\section{Comparison to Observations}
\label{comparison_with_observations}

\begin{figure}
\centering
\includegraphics[width=\hsize]{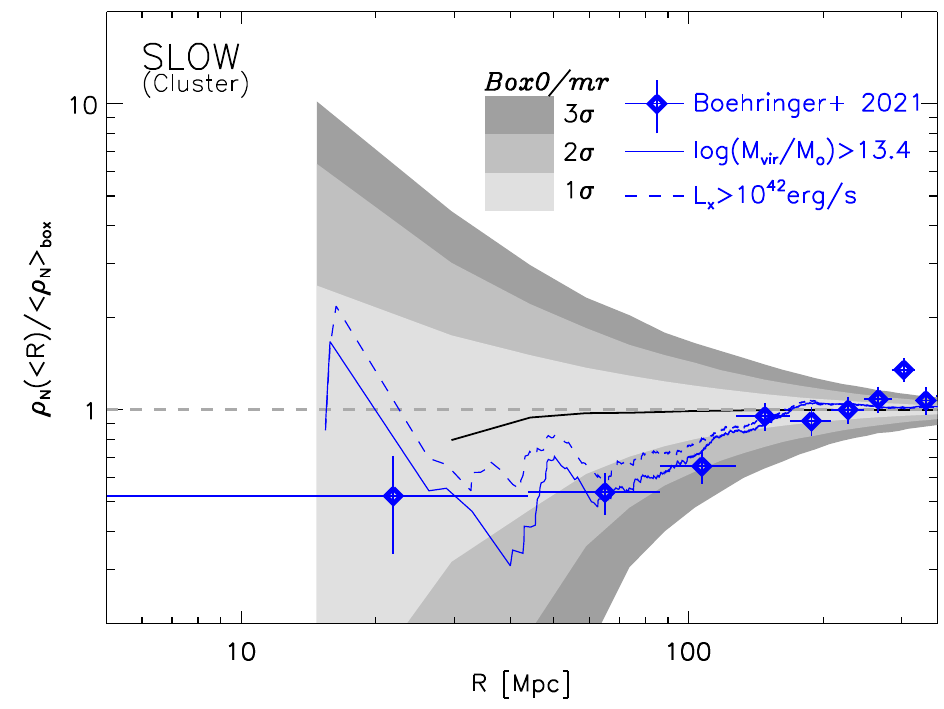}\\
\includegraphics[width=\hsize]{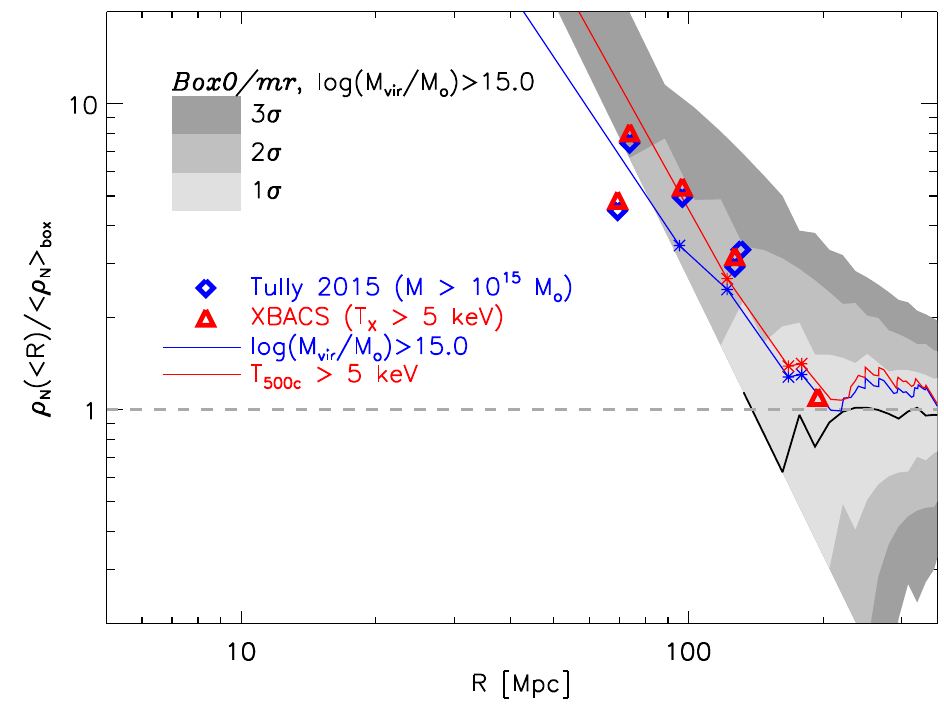}\\
\includegraphics[width=\hsize]{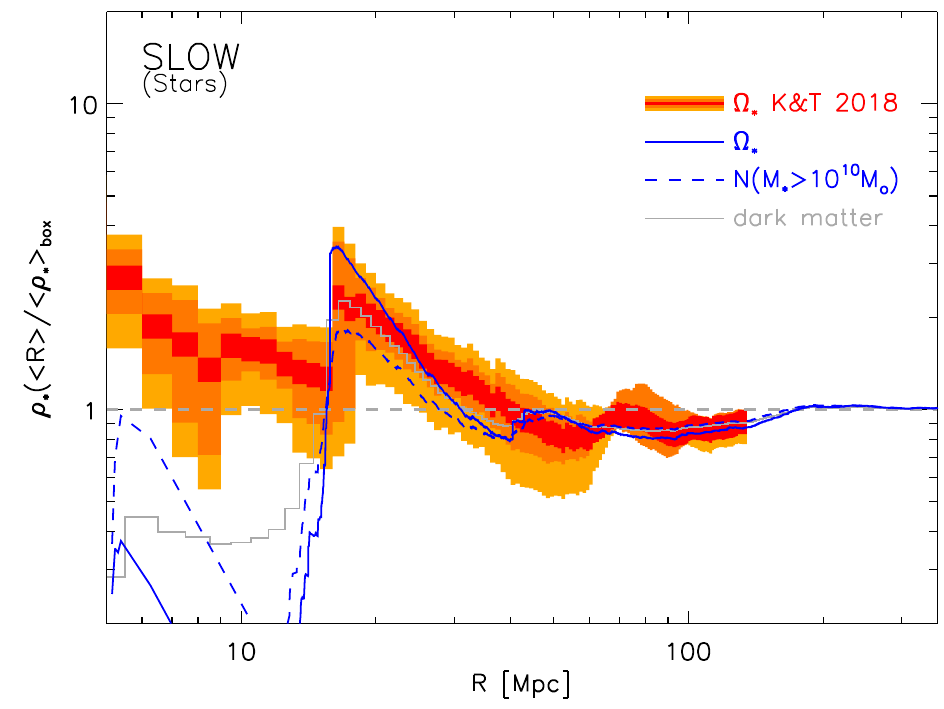}
\caption{Cumulative relative density of halos/galaxies with different masses as a function of distance (computed from the redshift) compared to different observations. Top panel comparing galaxy clusters to the X-ray sample from \citet{2021A&A...651A..15B}, using the same mass (solid) and the same X-ray luminosity (dashed line) cut. Middle panel, comparing massive galaxy clusters (M$_{\rm vir}>10^{15}$ blue line, T$_{500c} > 5$ keV red line) to the five closest clusters exceeding this mass or temperature from the Tully catalogue \citep{2015AJ....149..171T} or the BAX database \cite{2004A&A...424.1097S}, respectively. Bottom panel, comparing our simulated galaxies with the results on the stellar mass density presented in \citet{2018AN....339..615K}. In the upper two panels, the gray shaded regions mark the according 1,2 and 3 $\sigma$ lines obtained from the {\it Magneticum} simulation and the black line marks the median (to display the cosmic variance left).}
\label{DensityComparison}
\end{figure}

As discussed in the previous section, the different features in the Local Universe can be compared more directly to observations to understand better how unique our Local Universe is. Figure~\ref{DensityComparison} shows the comparison to different observational tracers as discussed in the following sub-sections in detail.

\subsection{X-ray cluster sample}

The upper panel of figure~\ref{DensityComparison} compares the results from the SLOW simulation to the findings by \citet{2020A&A...633A..19B}, who reported a 50\% under-density of X-ray selected galaxy cluster from the CLASSIX catalogue (shown as blue data points). The luminosity cut of $10^{42}$erg/s in the (0.1-2.4) keV band translates into a virial mass of $\approx 10^{13.4}\mathrm{M}_\odot$ when using a canonical X-ray luminosity-mass relation \citep{2014A&A...570A..31B}. The blue solid line resembles the result from the SLOW simulation for halos with this mass threshold, nicely reproducing the data points. However, having a full hydro-dynamical simulation we can also directly use the predicted X-ray luminosity of the clusters in the simulations. Here, we started from the predicted bolometric luminosity within $R_{500c}$ and applied the correction for the used energy band based on the mass weighted $T_{500c}$ of the cluster. When applying the X-ray luminosity cut to the simulated clusters, we get the dashed line, which is very close to the one where we used the virial mass cut to select the clusters. This shows that the observed signal in the simulation -- and therefore also the signal in the real data -- is not driven by the X-ray selection of galaxy clusters.

Given this signal, it is interesting to investigate how peculiar it is to have a local environment which features such a 50\% under-density within the given volume. Therefore, we took the very large, general cosmological simulation from the {\it Magneticum} simulations set, namely {\it Box0/mr}, which covers a volume of ($2688h^{-1}$Mpc)$^3$. Here we randomly selected more than fifteen thousand points within the volume, and computed the cumulative over-density profiles out to 360 Mpc radius. The gray shaded regions mark the one, two and three sigma regions occupied by these profiles from a random cosmological simulation. The black solid line is the median of the distribution to indicate the statistical error left due to the still somewhat limited sample size from the large simulation. We clearly see that the under-density we live in is not very uncommon in the cosmological sense, representing a $\approx$1.5 sigma event, similar to what \citet{2016MNRAS.455.2078S} concluded from comparing the halo mass function. 

\subsection{Massive galaxy clusters}

Although we are living in a large scale under-dense region, there are several very massive galaxy clusters within that region, exceeding virial masses of $10^{15}\mathrm{M}_\odot$, among them Coma, Perseus and Ophiuchus. Using the five closest galaxy clusters from the Tully galaxy groups catalogue \citep{2015AJ....149..171T} with masses above $10^{15}M_\odot$, the middle part of figure \ref{DensityComparison} shows that this corresponds to a very large over-density in the Local Universe (blue diamonds). We also selected the five closest galaxy clusters from the BAX\footnote{http://bax.irap.omp.eu and references therein.} database with X-ray measured temperatures exceeding 5 keV and showing them as red triangles in the figure, basically confirming the presence of an overdensity of high mass systems. Selecting such clusters predicted by the SLOW simulation (blue line for the mass selection, red line for the temperature selection) again follows the observational data points extremely well, confirming this significant overdensity of massive clusters, as it was already clearly visible in figure~\ref{DensityHaloes}. Note that there is a large overlap between the selected clusters when switching the selection criteria in both, observations (4/5) and simulations (3/5). Note that the closest cluster in the simulation exceeding $10^{15}M_\odot$ is  at a distance of $\approx 40$Mpc (as can be seen from figure \ref{DensityHaloes}). Furthermore, many of these prominent, massive/hot clusters can be cross identified between the simulations and observations, among them Perseus, Coma, A119 and A85, even show very similar virial masses and temperatures \citep[see][and for more details Hern\'andez-Mart\'inez et al. in prep]{2018MNRAS.478.5199S,2023arXiv230101305S}. Only Norma and Ophiuchus are not very well reproduced in the simulations, due to their position in/close to the zone of avoidance, where no data are available. Nevertheless, there is a significant overlap between the two sets of most massive / hot clusters in simulations and observations. Therefore we expect similar properties of the clusters in both sets as was already shown for example for the Virgo cluster \citep{2021MNRAS.504.2998S}.

Again we investigate the statistical significance of such excess of massive systems. Here, the gray shaded regions again mark the one, two and three sigma regions as obtained from the large cosmological simulation {\it Box0/mr} from the {\it Magneticum} set. The comparison indicates that such excess in massive system is approximately a 1.5 sigma coincidence. The black line marks the median from the 15635 samples we used and gives an indication on the remaining statistical uncertainty, which in this case is somewhat larger due to the general low number density of such massive galaxy clusters.

One might speculate, that the general under-density of clusters might be related to the over-density of massive systems, however, a closer investigation based on {\it Box0/mr} indicates the opposite. In fact, when requiring a very conservative limit to have at 90 Mpc a mean density below 0.65 for clusters with viral masses of $\approx 10^{13.4}M_\odot$ and having an over-density of massive systems of 2.3 at 160 Mpc (corresponding to the observational value of the the closest system), we find only 44 out of the 15635 samples to match. This corresponds to a more than $3\sigma$ case. Besides, it clearly demonstrates that these two peculiarities of the local Universe are not related.

It is also worth noticing that in addition, at distances between $\approx(240-330)$Mpc, the simulation predict a $\sim(20-30)$\% over-density of massive galaxy clusters, as visible in the middle panel of figure \ref{DensityComparison} as well as in the left panel of \ref{DensityHaloes}. Also at these distances, which are beyond the radial distance for which peculiar velocities are mainly constrained, some of the very massive galaxy clusters can be cross matched between simulations and observations, among them A3266 for example, which is also labeled in figure \ref{FigFullSky_obs} and \ref{FigFullSky_sim}. 

\subsection{Galaxies}
\begin{figure}
\centering
\includegraphics[width=\hsize]{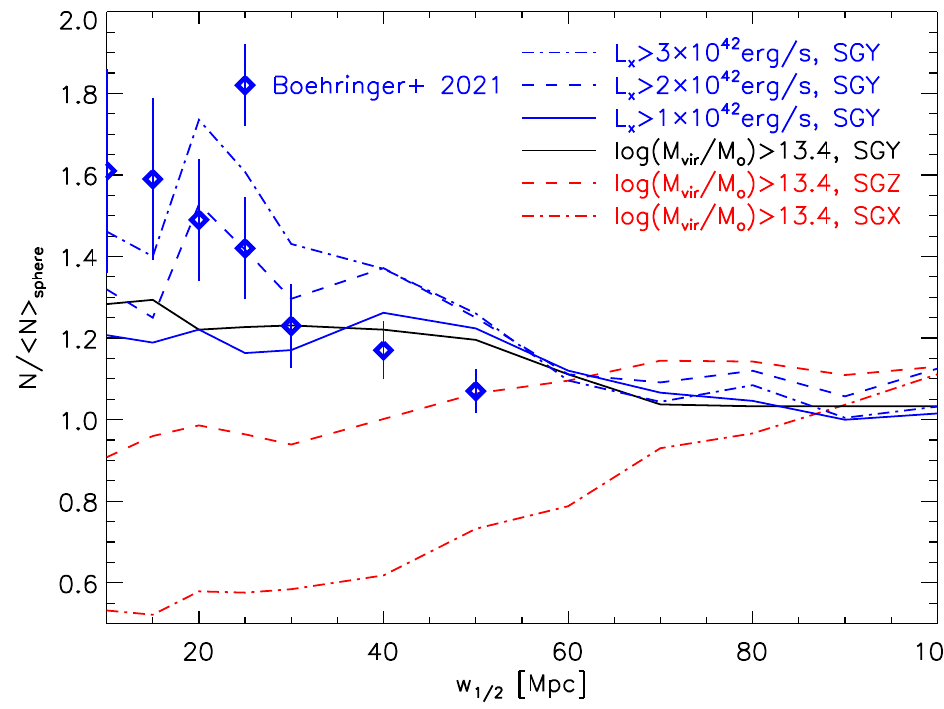}
\caption{Comparing the flattening of the galaxy clusters distribution to the X-ray sample from \citet{2021A&A...651A..15B} (see text for description), to the simulation, using the same mass (black) and the same X-ray luminosity (blue solid line) cut. The dashed and the dashed dotted red lines are applying the same measurement in the other, super galactic coordinates for comparison. The dashed and dotted dashed blue lines are slightly different cuts in X-ray luminosity.}
\label{AnisotropyComparisonCluster}
\end{figure}

Finally we can also compare the density fields as traced with galaxies, as this might be closer to the actual, underlying distribution of dark matter, as shown in the previous section. Here, in the lower panel of figure \ref{DensityComparison} we compare the stellar mass in all simulated galaxies in SLOW (solid blue line) with observed stellar density as reported in \citep{2018AN....339..615K}. Here the red band correspond to the overall mean stellar density reported in \citet{2018AN....339..615K}, normalized to the global stellar density of $\Omega_*= 0.0027$ \citep{2004ApJ...616..643F} or to  the value at the largest distance. The orange and the dark orange bands correspond to the observed stellar density for the northern and southern \citep{2018AN....339..615K}, normalized to these two general values respectively. In addition, the blue dashed line marks the relative, mean number density of galaxies with $M_*>10^{10}M_\odot$, while the gray line is the relative dark matter density. Comparing to the expectations from the SLOW simulation in both cases matched the large over-density feature of factor $\approx 2$ of the local structures associated with the Virgo cluster at distances around $\approx 16$ Mpc quite well. It also shows a very similar shape further out with two minima and two maxima, although the first under-density in the simulation appears at a distance of $\approx 40$ Mpc, while in the observations the minimum in the relative density is at $\approx 55$ Mpc. As seen in comparison with the dark matter density, these features are clearly directly related to the underlying dark matter density field. Closer than the distance to Virgo, the simulations show a very prominent, large under-density of the local void (already visible in figure \ref{DensityHaloes}, however, the stellar density as reported in \citet{2018AN....339..615K} features a larger, local density. Interestingly to note, the SLOW simulation predict a global stellar density of $\Omega_* = 0.0031$, which is close to the value of $\Omega_*= 0.0027$ as reported in \citet{2004ApJ...616..643F}. In general, the good match of the gray line for the dark matter density in the bottom panel with the observed as well as the simulated stellar density at distances beyond $\approx 20$ Mpc indicates that the observed stellar densities obtained from galaxies quite robustly trace the underlying dark matter field, which is not the case when using galaxy clusters as tracers, as shown in the previous sub-sections. 

\subsection{Anisotropies}

\begin{figure}
\centering
\includegraphics[width=\hsize]{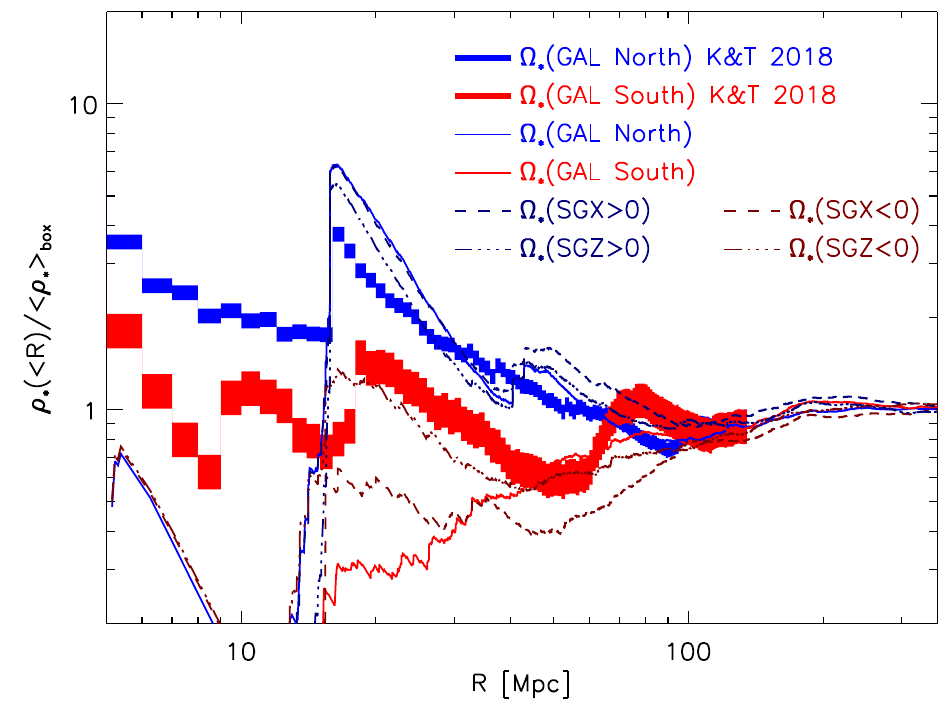}
\caption{Comparison of our simulated galaxies with the results on the stellar mass density presented in \citet{2018AN....339..615K} divided in the Northern Galactic hemisphere (blue) and Southern Galactic hemisphere (red). As comparison, the thinner, dark red dashed and dotted dashed line show the result when splitting along super galactic SGX and SGZ coordinates. }
\label{AnisotropyComparisonGalaxies}
\end{figure}

As mentioned in the introduction, the local structures also lead to observed non isotropic distribution of tracers of the large scale structures. Here we want to compare the results of the SLOW simulations with two of these reported features. In figure \ref{AnisotropyComparisonCluster} we compare the distribution of galaxy clusters in the simulations to the pancake like structure out to a scale of $\approx$ 100 Mpc as reported in \citet{2021A&A...651A..15B}. We follow the characterization of the flattened superstructure as in the observations. Starting from a cylinder with 100 Mpc radius, centered at the MW position and oriented along the supergalactic SGY direction, we compute the over-density within slices of thickness $\pm w_{1/2}$ Mpc and compare this with the mean density of the same objects within a sphere of 100 Mpc radius, centred on the MW. To mimic the observational selection of clusters we selected clusters either by their X-ray luminosity (e.g. $10^{42}$erg/s as in the observations) or by the according virial mass (e.g. $M_\mathrm{vir}=10^{13.4}M_\odot$) as shown by the blue and black solid line respectively. This qualitatively agrees well with the observational data points (blue symbols with error bars) which are also shown. In addition, the blue dashed and dashed-dotted lines show the dependence on the luminosity threshold used, indicating that choosing a slightly larger luminosity would increase the agreement with the observational data further. To further strengthen this result, the red dashed and dotted-dashed lines are showing the absence (or even reverse signal) obtained when doing the split along the supergalactic SGZ and SGX coordinate, respectively. We can conclude here, that the pancake like structure is a solid prediction by the SLOW simulation and well aligns qualitatively with the observational finding without fine tuning selection parameters. 

In figure \ref{AnisotropyComparisonGalaxies}, we repeat the comparison of the stellar density within the Local Universe with the finding of \citet{2018AN....339..615K}, where a large differences between the Northern and Southern Galactic hemisphere was reported. Splitting our galaxies from the SLOW simulation in the same way (red and blue solid line) the SLOW simulation shows some features very similar to the observations (red and blue bands). Clearly, there is an over respectively under density present in the northern respectively southern hemisphere at large distances. This qualitatively agrees with observations, although here the quantitative agreement is not as good as for the pancake like structure traced by galaxy clusters. Also here we add in addition a split along the super-galactic x and z coordinate (dashed and dashed-dotted lines) to show the dependence of the signal onto the directional split. As seen before, the lower stellar density within the SLOW simulation out to a distance of $\approx15$ Mpc is present and independent on the direction. Interestingly the split along the super-galactic x coordinate follows the observations even better and may indicate that the positional uncertainty of some prominent structure in the simulated Local Universe might influence the actual selection and might be worth further studies.

\section{Conclusions}
\label{conclusions}

We presented the first results from a long standing initiative, to perform hydrodynamical, cosmological simulations of the Local Universe (a) with high enough resolution and detailed enough galaxy formation physics to reproduce reliable galaxies as well as intracluster medium properties; (b) which extends to large distances to properly cover the transition of the local structures towards the cosmological average properties; (c) to be constructed not on total density estimates based on galaxy densities in redshift surveys. The latter makes the comparisons with the observed galaxy distribution and galaxy clusters properties independent from observations, that where already used to construct the constraints for the initial condition and also allow to better compare their evolution. Initial conditions of our SLOW simulation are based on peculiar velocities (more precisely, a complex interplay and combination of observed redshift and distance modulus from the CosmicFlows-2 catalogue), applying various improvements in their creation over the last years \citep{2018MNRAS.476.4362S,2020MNRAS.495.4463S} leading to a simulation which captures a volume of (500 $h^{-1}$Mpc)$^3$ in which various clusters of the Local Universe can be cross identified (see \citet{2018MNRAS.478.5199S} and Hern\'andez-Mart\'inez et al, in prep, for details).

The predicted density field of the simulation shows various distinct features and indicates that the Local Universe transits into the cosmic mean at distance of $\approx$ 200 Mpc (with significant variances depending on the tracers used).
Within this region, several ranges with relative under- and over-density are present, which can be compared to observational indications. Especially we find that  
\begin{enumerate}
\item Up to the distance of Virgo cluster (i.e. $\approx16$ Mpc), the SLOW simulation predicts the mean dark matter density in the Local Universe to be at $\approx$ 0.5 of the mean value. With the Virgo super cluster this transits into a factor of two over-density region until $\approx$ 30 Mpc, after which the Local Universe seems to be $\approx$ 20\% under-dense out to $\approx$ 200 Mpc.  
\item While this is traced by normal galaxies within the simulation with an expected bias of order of 30\%, using halos or galaxy clusters can result in quite different, sometimes even opposite conclusions, depending on the halo mass cut.
\item Applying the same mass (or alternatively X-ray luminosity) cut like the CLASSIX galaxy cluster sample, the simulation very closely reproduces the observed 50\% under-density of galaxy clusters in the Local Universe.
\item Using clusters with virial masses above $10^{15}M_\odot$, simulation and observations consistently show a significant over-density of such objects within the same volume out to $\approx 200$ Mpc within the Local Universe. In addition, at distances between $\approx(240-330)$~Mpc, the simulation predicts a $\approx(20-30)$\% over-density of such massive galaxy clusters.
\item Comparing with hydro-dynamical simulations of very large cosmological volumes, we find that both these features individually are not such uncommon and correspond to $\approx 18$\% of cases (e.g. fall into the $\approx$1.5 sigma region).
However, they appear to be unrelated to each other and the combination is only found in $\approx0.28$\% of random selections from our (4 Gpc)$^3$ reference simulation {\it Box0/mr} of the {\it Magneticum} simulation set, and therefore would correspond to a three sigma case).
\item The SLOW simulation also shows the pancake-like distribution of galaxy clusters within 100 Mpc of the Local Universe \citet{2021A&A...651A..15B}. Thereby the radial distribution of the number counts of galaxy clusters follows quite closely the observations when splitting in super-galactic north/south direction.
\item The SLOW simulation predicts a global stellar density of $\Omega_* = 0.0031$, which is close to the value of $\Omega_*= 0.0027$ as reported in \citet{2004ApJ...616..643F}. At distances larger than $\approx 15$ Mpc the stellar density obtained from the SLOW simulation follows the one reported in \citet{2018AN....339..615K} remarkably well and confirms both, the large, relative over-density induced by Virgo as well as the $\approx 20$\% under-density beyond distances of 100 Mpc.
\item At distances between $\approx 15$ and $100$ Mpc, the predicted stellar density in the slow SLOW simulation shows also qualitatively similar difference between the northern and southern Galactic hemisphere than reported in \citet{2018AN....339..615K}.

\end{enumerate}

The presented SLOW simulation of the Local Universe reproduces some of the main features of the local density field and therefore opens a new window for local field cosmology. It allows us to better verify and interpret observations of the local structures and their tracers. In future this will allow us to evaluate in detail the imprint of the specific density field of the Local Universe on the local estimations of cosmological parameters like H$_0$ as well detailed studies of the imprint of the formation history on actual properties of galaxy clusters.

\begin{acknowledgements}
%\section*{Acknowledgements}
We want to thank T. Hoffmann for the help when extracting observational data points from the according publication and McAlpine for providing the mean dark matter density profile of the Sibelius simulation, which we added to complete the figures in the appendix. This work was supported by the grant agreements ANR-21-CE31-0019 / 490702358 from the French Agence Nationale de la Recherche / DFG for the LOCALIZATION project. NA acknowledges support from the European Union’s Horizon 2020 research and innovation program grant agreement ERC-2015-AdG 695561. KD and MV acknowledge support by the Excellence Cluster ORIGINS which is funded by the Deutsche Forschungsgemeinschaft (DFG, German Research Foundation) under Germany’s Excellence  Strategy – EXC-2094 – 390783311 and funding for the COMPLEX project from the European Research Council (ERC) under the European Union’s Horizon 2020 research and innovation program grant agreement ERC-2019-AdG 882679. MV also acknowledges support from the Alexander von Humboldt Stiftung and the Carl Friedrich von Siemens Stiftung. The calculations for the hydro-dynamical simulations were carried out at the Leibniz Supercomputer Center (LRZ) under the project pr83li (Magneticum) and pn68na (SLOW). We are especially grateful for the support by M. Petkova through the Computational Center for Particle and Astrophysics (C2PAP). 
\end{acknowledgements}

\bibliographystyle{aa}
\bibliography{bibfile}

\begin{appendix} %First appendix

\section{Comparison to other simulations}
\label{comparison_sims}

Here we present some comparison with other, previous or more recent constrained simulations, like CORUSCANT \citep{2004JETPL..79..583D} and its SALACIOUS variant \citep{2007ApJ...659..257K} which includes galaxy formation physics, the early CLUES \citep{2003ApJ...596...19K} simulation (label as B160\_WM3) as well as the SIBELIUS \citep{2022MNRAS.512.5823M} simulation.

\begin{figure}
\centering
\includegraphics[width=0.99\hsize]{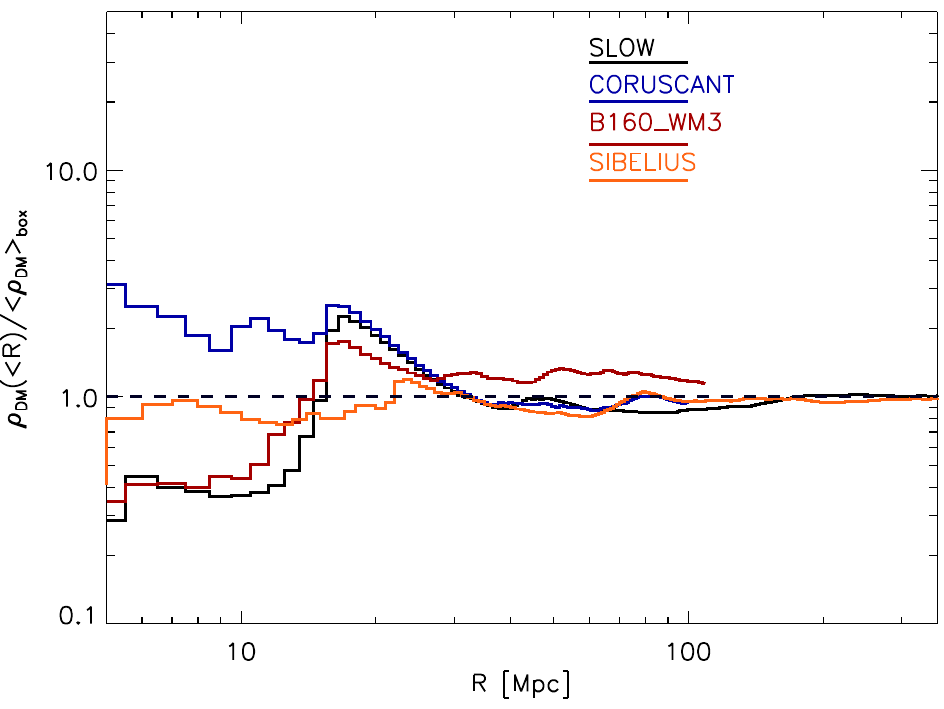}
\caption{Cumulative, relative dark matter density as function of distance from the MW position as obtained from the SLOW (black), CORUSCANT (blue), early CLUES (red) and SIBELIUS (orange) simulation.}
\label{DensitySims1}
\end{figure}

\subsection{Dark matter distribution}

Similar to figure \ref{DensityHaloes}, figure \ref{DensitySims1} shows the dark matter density in the Local Universe but comparing different simulation: SLOW, CORUSCANT, B160\_WM3 and SIBELIUS. Although for SIBELIUS the dark matter data are not publicly available, the dark matter profiles where provided by McAlpine on our request. All simulations show the over density related to the Virgo complex, however, SIBELIUS shows only a very mild signal. On the other hand, SIBELIUS shows a significant  under density at scales of 30-60 Mpc, while only SLOW shows the large scale under density towards a distance of 100 Mpc. For B160\_WM3 and CORUSCANT this is mainly because the volume of these simulations cover a too small volume, while there is a very tiny under density visible in SIBELIUS. However, B160\_WM3 shows a significant over density within a spherical region of 100 Mpc radius. Interestingly, both, SLOW and B160\_WM3 show a clear local under density in dark matter within 10 Mpc, while CORUSCANT shows a larger over density in this region. 

\subsection{Galaxy cluster distribution}

\begin{figure}
\centering
\includegraphics[width=0.99\hsize]{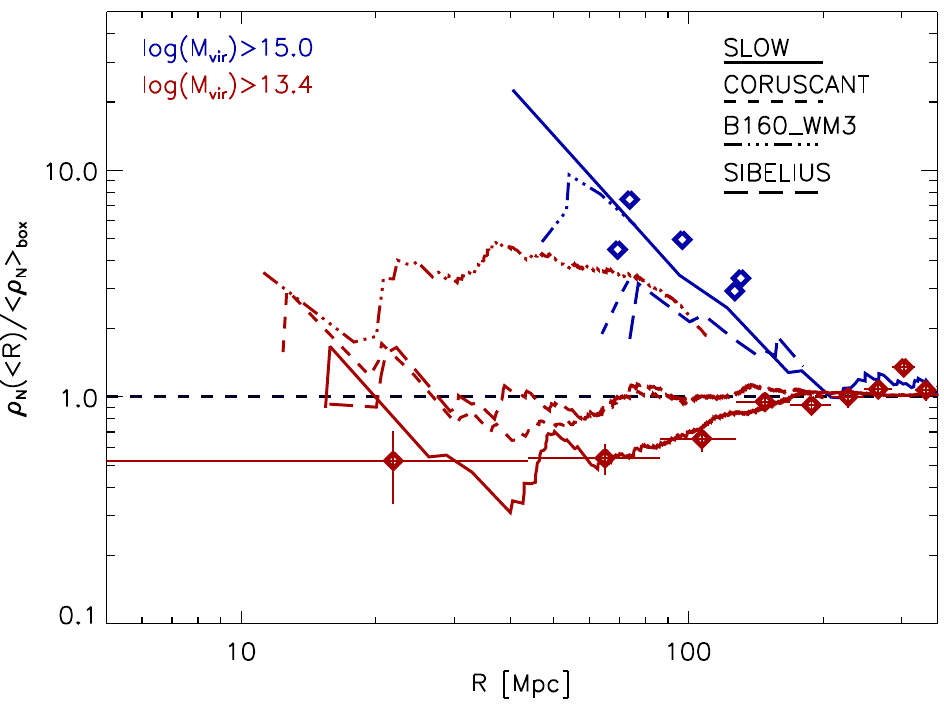}
\caption{Equivalent to top and middle panel of figure \ref{DensityComparison} but for the different simulations, showing the cumulative, relative cluster density as function of distance from the MW position as obtained from the SLOW (solid), CORUSCANT (dashed), SIBELIUS (long dashed) and early CLUES (dashed dotted) simulation. The blue lines are for very massive halos, the red line are for clusters and correspond to the observations from the CLASSIX \citet{2020A&A...633A..19B} sample.}
\label{DensitySims2}
\end{figure}

Similar to figure \ref{DensityComparison}, in figure \ref{DensitySims2} we show the density of galaxy clusters in the Local Universe as obtained from the different simulations, divided into massive (blue) and all clusters (red), compared to the observational data points. While in SLOW and B160\_WM3 reproducing the high number of local, very massive system, CORUSCANT and SIBELIUS are falling somewhat short in having the unusual large number of such massive systems. Note however that this is not really significant, given the low number of halos in this case. However, when comparing to the CLASSIX sample, only SLOW is able to match the reported significant under density out to a distance of 100 Mpc.  

\subsection{Galaxy density profiles}

\begin{figure*}
\centering
\includegraphics[width=0.45\hsize]{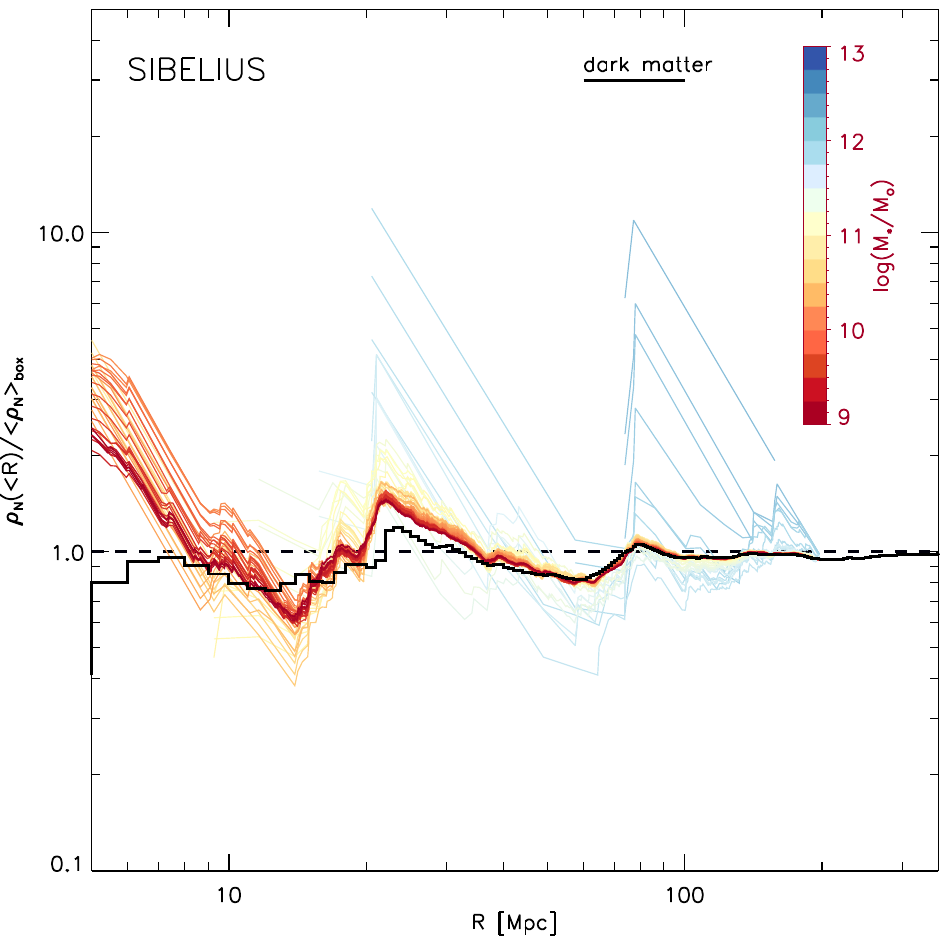}
\includegraphics[width=0.45\hsize]{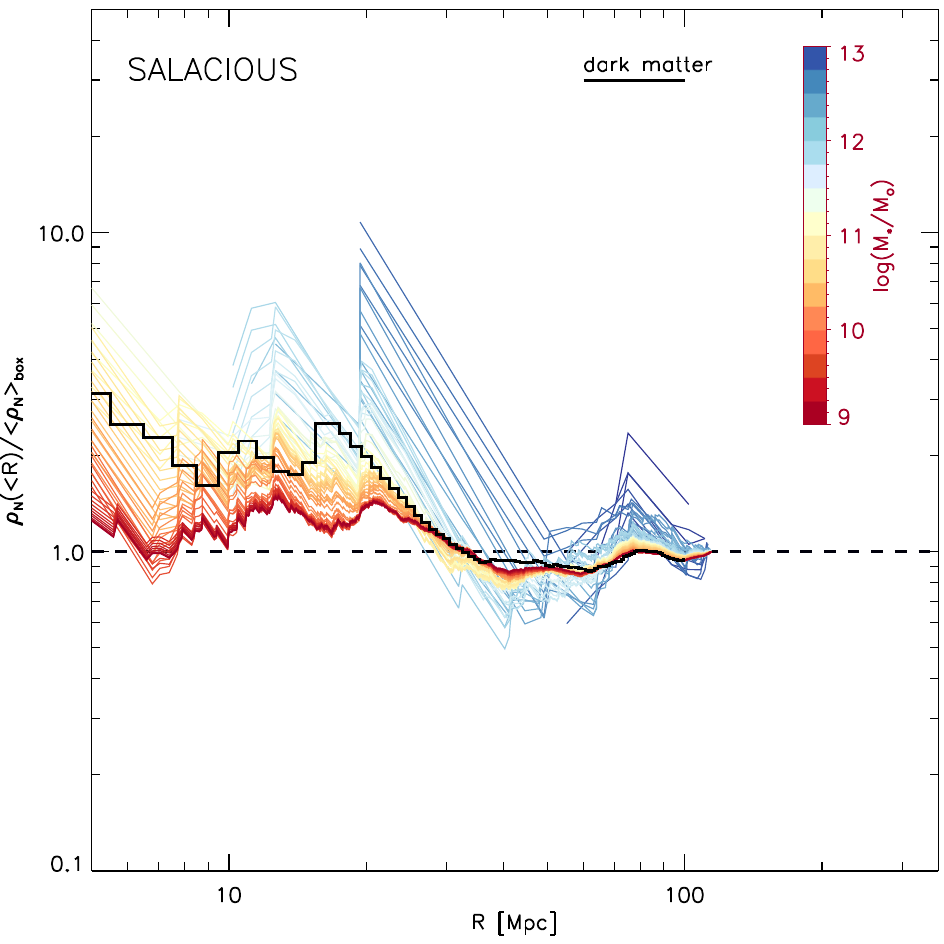}
\caption{Same as right panel of figure \ref{DensityHaloes} but for SIBELIUS (left panel) and SALACIOUS (right panel). Shown is the mean density of galaxies within the Local Universe for different stellar masses (as indicated in the color bar). Additionally the dark matter density is also shown in both cases.}
\label{DensitySims3}
\end{figure*}

Finally we repeat the right panel of figure \ref{DensityHaloes} for SIBELIUS (left panel) and SALACIOUS (right panel) in figure \ref{DensitySims3}. Shown is (as before) the mean density of galaxies as function of distance for different stellar masses, as indicated in the color bar. Note, that from this we obtain a  $\Omega_* = 0.0014$ for SIBELIUS and $\Omega_* = 0.0021$ for CORUSCANT, compared to $\Omega_* = 0.0031$ for SLOW and the observed value of $\Omega_*= 0.0027$ as reported in \citet{2004ApJ...616..643F}. Compared to the SLOW simulations, both simulations are displaying significant less of variation in the mean density, and especially the SIBELIUS does not show one sided deviation from the mean density over very large scales. As already discussed in figure \ref{DensitySims2}, SIBELIUS does not show the large scale under density out to a distance of 100 Mpc, as reported in the CLASSIX sample of galaxy clusters. Here it gets clear, that contrary to SLOW, no galaxy or galaxy cluster selection in SIBELIUS does feature this observed anomaly in the Local Universe. Once more this demonstrates that the assumption of a constant bias within reconstructions based on galaxy densities limits the predictive power of the resulting constrained simulations. The significant differences of the bias and its environmental dependency visible in simulations based on semi-analytic modeling (like SIBELIUS) and the full hydrodynamical simulations (like SLOW or CORUSCANT) emphasize that this is a non-trivial obstacle for reconstructions of the local Universe. This can in principle be overcome with reconstructions based on the observed velocity field, as demonstrated through the SLOW simulation, which seems to match the observational findings of various different tracers of the large scale structure very well. However, as shown, they come with their own obstacles which are difficult to overcome in a satisfactory manner, as demonstrated by the overall effort which has been spent to bring forward the constrained simulations to this point. 

\subsection{Weighting haloes by mass}
\begin{figure}
\centering
\includegraphics[width=0.99\hsize]{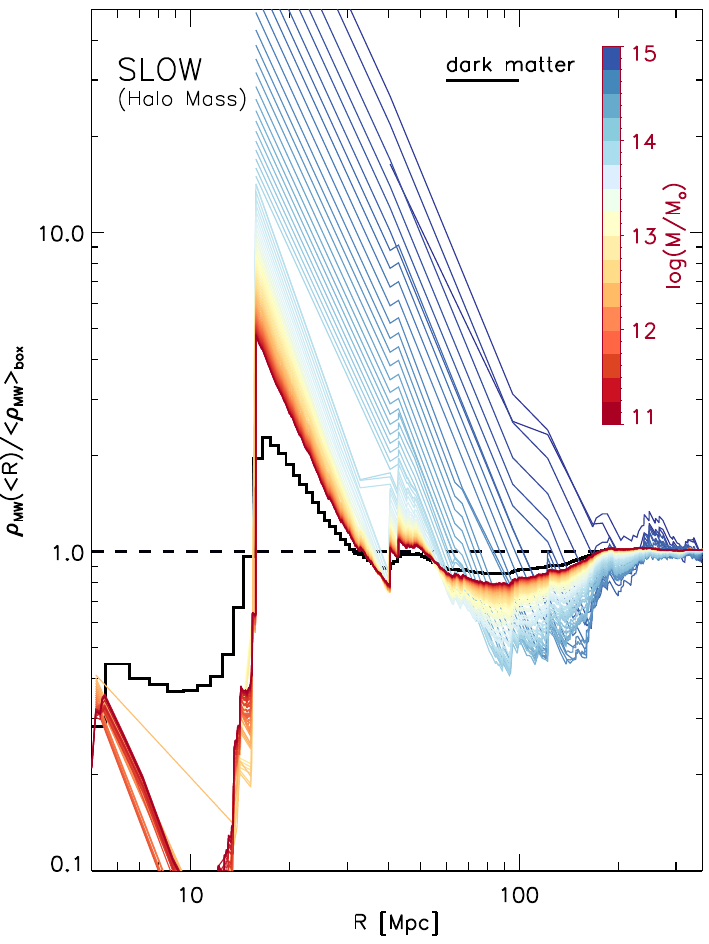}
\caption{Same as left panel of figure \ref{DensityHaloes}, but using mass weighting to compute the density, instead of number density. }
\label{MassDensityHaloes}
\end{figure}

We also tested if the mass weighting of haloes would result in densities closer to the dark matter density, especially for the peak created by the Virgo cluster region, where one could think that in the case of using  groups and clusters, this might give a better result. The result is shown in figure \ref{MassDensityHaloes}, which is identical to the left part of figure \ref{DensityHaloes}, except that we used mass weighting to compute the densities instead of number densities. Especially for the nearby structures, like the local void and the peak dominated by the Virgo cluster region, the mass weighting even overshoots the feature. However, the mass weighting creates a very sharp feature at a distance of $\approx 40$ Mpc, even when using down to very small haloes, which is similar to the sharp feature visible in the stellar density at distance of $\approx 30$ Mpc found by \citet{2018AN....339..615K} (see lower panel of figure \ref{comparison_with_observations}), which is obtained by integrating the observed stellar mass function.

\end{appendix}

\end{document}